\documentclass[
    11pt,
    letterpaper,
    reprint,
    notitlepage,
    superscriptaddress,
    aps,prx
]{revtex4-2}

\usepackage{amsmath,amssymb,amsfonts} 
\usepackage{empheq}
\usepackage{physics}
\usepackage{lipsum}

\usepackage[normalem]{ulem}

\usepackage{microtype}

\usepackage{bm}

\usepackage[svgnames,dvipsnames]{xcolor}
\usepackage{graphicx}
\definecolor{NewBlue}{rgb}{0.1, 0.1, 0.7}
\definecolor{NewRed}{rgb}{0.7, 0.1, 0.1}
\usepackage[colorlinks,
    linkcolor=Maroon,
    citecolor=NewBlue,
    urlcolor=ForestGreen]{hyperref}

\usepackage[capitalise]{cleveref}

\usepackage[
    exponent-product=\cdot,
    range-phrase=\text{--},
    range-units=single,
    separate-uncertainty,
]{siunitx}[=v2]

\newif\ifwritetoc
\writetocfalse

\let\origaddcontentsline\addcontentsline
\renewcommand{\addcontentsline}[3]{%
  \ifwritetoc
    \origaddcontentsline{#1}{#2}{#3}%
  \fi
}

\renewcommand{\t}[1]{\mathrm{{#1}}}

\newcommand{\dop}[1]{\delta\hat{#1}}

\newcommand{\eqdef}{\coloneqq}

\newcommand{\NTT}{Physics \& Informatics Laboratories, NTT Research, Inc., Sunnyvale, California 94085, USA}
\newcommand{\LigoMIT}{LIGO Laboratory, Massachusetts Institute of Technology, Cambridge, MA 02139, USA}
\newcommand{\MechMIT}{Department of Mechanical Engineering, Massachusetts Institute of Technology, 
    Cambridge, MA 02139, USA}
\newcommand{\Stanford}{Edward L. Ginzton Laboratory, Stanford University, Stanford, California 94305, USA}

\begin{document}

\title{Continuum-field quantum optics of frequency comb metrology}

\author{Dong-Chel Shin}
\email{dongchel@mit.edu}
\affiliation{\MechMIT}
\affiliation{\NTT}

\author{Edwin Ng}
\email{edwin.ng@ntt-research.com}
\affiliation{\NTT}
\affiliation{\Stanford}

\author{Myoung-Gyun Suh}
\email{myoung-gyun.suh@ntt-research.com}
\affiliation{\NTT}

\author{Vivishek Sudhir}
\email{vivishek@mit.edu}
\affiliation{\MechMIT}
\affiliation{\LigoMIT}

\date{\today}

\begin{abstract}
Frequency combs enable precision measurements across timekeeping, spectroscopy, ranging and astronomy, and are now extending to integrated and field-deployable platforms. 
Realizing their full performance demands a comprehensive account of the quantum noise that arises when broadband optical fields are converted into finite-bandwidth electrical signals. 
Here we present a rigorous first-principles quantum-mechanical framework for optical frequency-comb metrology based on continuous-mode field quantization and a comb-line-resolved description of quantum fluctuations. 
The theory describes how quantum fluctuations of the comb field are transduced into electrical measurement noise.
Formulated at the level of the comb field, the framework unifies the standard quantum limits of optical frequency division (OFD) and dual-comb spectroscopy (DCS) within a single treatment, and provides a general recipe for other comb-based measurements.
On this footing, we identify practical, resource-efficient routes to quantum enhancement through engineered comb states, laying a foundation for the design of next-generation frequency combs operating at and beyond standard quantum limits.
\end{abstract}

\maketitle

\section{Introduction}
Optical frequency combs establish a direct, phase-coherent link between optical and microwave frequencies~\cite{Udem99,Jones00,Diddams00,Udem02,diddams20}.
Their spectrum---a set of discrete, evenly spaced lines whose frequencies $\Omega_c + n\Omega_r$ are fixed by just two radio-frequency parameters, the carrier-envelope offset $\Omega_c$ and the repetition rate $\Omega_r$---acts as a precise ruler for optical frequencies.
This simple yet rigid structure has enabled optical atomic clocks with fractional instabilities below $10^{-18}$~\cite{Diddams01,CundiffYe03}, broadband molecular fingerprinting~\cite{Coddington08,Ycas18,picque19}, sub-micrometer ranging~\cite{Coddington09}, and astronomical spectrograph calibration for exoplanet searches~\cite{Steinmetz08,Suh19}, capabilities now extending to compact, fieldable platforms through microcombs~\cite{Kippenberg18}.

Every comb-based measurement terminates in broadband photodetection, mapping the optical field to a finite-bandwidth radio-frequency (RF) or microwave signal.
As classical technical noise is progressively suppressed, the residual photocurrent noise 
approaches the fundamental quantum noise of the optical field~\cite{Caldwell22,Caldwell23,herman25}: during photodetection, each comb line acts as a local oscillator that heterodynes the vacuum fluctuations in its spectral vicinity down to the RF domain~\cite{Quinlan2013JOSAB}.
Semiclassical treatments first revealed the role of this correlated photocurrent 
shot noise~\cite{Quinlan2013JOSAB,Newbury10,Walsh23}, 
but these did not treat the comb as a quantum field and so could not reveal strategies for
quantum engineering to suppress this noise. 
More recently, a discrete-mode quantum description of dual-comb 
spectroscopy ~\cite{Shi23,Hariri25,Zhuang25} showed that quantum resources like squeezing 
and entanglement can enhance sensitivity, and a measurement theory was formulated for comb interferometry with nonclassical light~\cite{Lordi24}. 
A unifying continuous-mode field description—of the kind that underpins single-mode quantum metrology~\cite{Blow90,DingSud24,LIGO2019,CasBow21,ligoFDS}—has, however, not yet been extended to combs, where photocurrent noise must be traced to vacuum fluctuations around each of thousands of comb lines rather than a single carrier.

\begin{figure}[t!]
    \centering
    \includegraphics[width=0.45\textwidth]{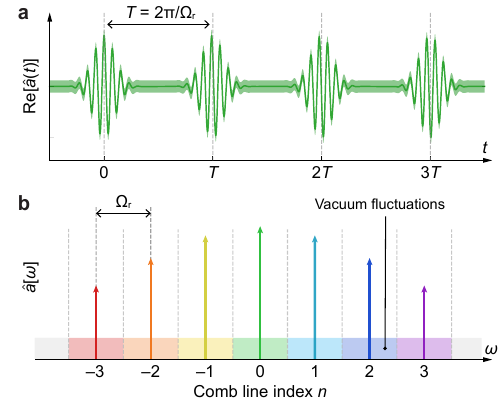}
    \caption{\textbf{Quantum model of a frequency comb.}
    (a)~Time-domain representation of the comb field: a train of coherent pulses with a uniform vacuum-noise band of constant width superimposed.
    (b)~Frequency-domain decomposition: coherent comb lines $\{\alpha_n\}$ sit atop independent bins of width $\Omega_r$, each associated with a comb-line-resolved operator $\delta\hat{a}_n[\Omega]$.}
    \label{fig:concept}
\end{figure}

Here, we develop such a framework from first principles---a continuum quantum theory of frequency comb metrology built on a comb-line-resolved decomposition of the quantum fluctuations of the comb field (Fig.~\ref{fig:concept})---and apply it to two canonical settings: optical frequency division (OFD)~\cite{Fortier11,Xie17,Quinlan2013NatPhot}, which transfers optical stability to the microwave domain, and dual-comb spectroscopy (DCS)~\cite{Coddington2016}, which encodes broadband optical spectra into RF interferograms.
For both, we derive closed-form linearized quantum noise expressions valid for arbitrary input states, reveal how the per-line quantum fluctuations map into the measurement noise, and identify specific quantum states that surpass the standard quantum limits.
The framework provides a general recipe for analyzing the quantum noise of any comb-based measurement, informing the design of next-generation quantum-enhanced frequency comb instruments.

\section{Results}

\subsection{Comb-line-resolved quantum field decomposition}

We model the frequency comb as a linearly-polarized electromagnetic field propagating 
along a fixed direction, with negligible dispersion and loss during propagation, such that
its classical part is an infinite periodic pulse train.
That is, classical part of its electric field is of the form:
$E(t) = \sum_n E_n e^{-i(\omega_0 + n\Omega_r) t}$, where 
$\omega_0$ is the (optical) carrier frequency and $\Omega_r$ is the repetition rate.
Clearly, the classical comb field consists of a discrete set of spectral lines at
frequencies $\omega_n = \omega_0 + n\Omega_r$, with complex amplitudes $\{E_n\}$ encoding 
the shape of the pulse.

The quantum description of the comb field is obtained by promoting the classical electric field
to a quantum operator and demanding that its expectation value reproduces the classical comb.
To wit, the continuous-mode quantization of the electromagnetic
field~\cite{Cohen97,Blow90} (see SI for details) yields the electric field operator
\begin{equation}
    \hat{E}(t) = \int_0^\infty \frac{d\omega}{2\pi}
    \mathcal{E}_\t{zp}(\omega)
    \left[i\hat{a}[\omega] e^{-i\omega t}+\text{H.c.}\right],
    \label{eq:E_field_main}
\end{equation}
where $\mathcal{E}_\t{zp}^2(\omega) = \hbar\omega/(2\mathcal{A}\epsilon_0 c)$, 
is the zero-point field variance per unit frequency when $\hat{E}$ is 
normalized to the optical power through the cross-section area $\mathcal{A}$. 
The operators $\hat{a},\hat{a}^\dagger$ obey the commutation relation
\begin{equation}
    [\hat{a}[\omega], \hat{a}[\omega']^\dagger] = 2\pi\,\delta[\omega - \omega'].
    \label{eq:comm_main}
\end{equation}
Requiring that the expectation value of $\hat{E}(t)$ reproduces the classical comb
field $E(t)$ is tantamount to the decomposition $\hat{a}[\omega] 
= \alpha[\omega] + \delta\hat{a}[\omega]$, where the coherent amplitude is 
$\alpha[\omega] = \sum_n \alpha_n 2\pi \delta[\omega-\omega_n]$,
with $\alpha_n = -iE_n/\mathcal{E}_\t{zp}(\omega_n)$,
and the quantum fluctuations are described by $\delta\hat{a}[\omega]$, which inherits 
the commutation relation in \cref{eq:comm_main}.

For the analysis of comb-based measurements, it is convenient to
adopt a description of the quantized field that resolves the quantum fluctuations around each classical comb line.
So consider the operator $\hat{a}_n[\Omega] = f[\Omega]\hat{a}[\omega_n + \Omega]$ where $f[\Omega]$ is a frequency-domain window centered at $\Omega=0$ and decaying thereafter, so that $\hat{a}_n$
represents the field around the comb line at $\omega_n$. 
The full field operator can be 
perfectly reconstructed from $\{\hat{a}_n\}$, as $\hat{a}[\omega]= \sum_n g[\omega-\omega_n] \hat{a}_n[\omega-\omega_n]$,
if the decomposition and reconstruction windows $f,g$ satisfy the condition $\sum_n g[\omega-\omega_n]f[\omega-\omega_n] = 1$ for all $\omega$.
Furthermore, the transformation $\hat{a}\mapsto \{\hat{a}_n\}$ is unitary if 
$[\hat{a}_n[\Omega], \hat{a}_m[\Omega']^\dagger] = 2\pi\,\delta_{nm}\,\delta[\Omega-\Omega']$; this requires that 
$\abs{f[\Omega]}^2 = 1$ and that $f$ is supported only on
$[-\Omega_r/2, \Omega_r/2]$. 
Smooth or non-compact windows, even when properly normalized, necessarily break unitarity; the unique valid choice is therefore a unit-height rectangular window of width $\Omega_r$, as is the reconstruction window $g$. (See SI Sec.~\ref{app:comb_line_resolved} for details and a more general discussion of the decomposition and reconstruction conditions.) 
Importantly, since the decomposition is unitary, the quantum states of the original field 
$\hat{a}[\omega]$ and the partitioned fields $\{\hat{a}_n[\Omega]\}$ are in 
one-to-one correspondence. 

Thus, we may think of the comb, in the frequency domain, 
as a collection of classical comb lines of amplitude $\alpha_n$ surrounded by 
quantum fluctuations associated with each line in a spectral bin of width $\Omega_r$ 
around it [see \cref{fig:concept}(b)], described by the 
operator $\delta\hat{a}_n[\Omega] = f[\Omega] \delta\hat{a}[\omega_n + \Omega]$.
These operators satisfy the commutation relation,
\begin{equation}\label{eq:comm_partitioned_main}
    [\delta\hat{a}_n[\Omega], \delta\hat{a}_m[\Omega']^\dagger]
    = 2\pi\,\delta_{nm}\,\delta[\Omega - \Omega'].
\end{equation}
These per-line quantum fluctuations clarify the origin of quantum noise
in comb-based measurements, and suggest strategies for quantum engineering to evade it.
For the description of the detection of the quantized comb field, it proves useful to move into
the time-domain:
\begin{align}
    \hat{a}(t)
    &= \int_0^\infty \frac{d\omega}{2\pi}\,\hat{a}[\omega]\,e^{-i\omega t} \notag\\
    &= \sum_n \left(\int_{-\Omega_r/2}^{\Omega_r/2}
        \frac{d\Omega}{2\pi}\,\hat{a}[\omega_n+\Omega]\,e^{-i\Omega t}
      \right)e^{-i\omega_n t} \notag\\
    &\equiv \sum_n \hat{a}_n(t)\,e^{-i\omega_n t}.
    \label{eq:partition_main}
\end{align}
Incorporating the classical amplitudes $\alpha_n$ gives the time-domain 
comb-line-resolved decomposition:
\begin{equation}
    \hat{a}(t) = \sum_n \left[\alpha_n + \delta\hat{a}_n(t)\right] e^{-i\omega_n t}.
    \label{eq:comb_decomp}
\end{equation}
In all comb-based measurements, the RF/microwave detection bandwidth $\Omega$ is
much smaller than the optical carrier frequencies ($\Omega \ll \omega_n$); in this
limit the broadband photoemission-rate operator~\cite{yurke85} reduces to the
instantaneous photon flux, yielding a photocurrent operator
\begin{equation}
\hat{I}(t) = q_e\,\hat{a}^\dagger(t)\hat{a}(t),
\end{equation}
where $q_e$ is the elementary charge (see SI Sec.~\ref{sec:broadband_SI}).
Linearizing around the coherent amplitudes then isolates a quantum noise term driven
entirely by $\{\delta\hat{a}_n(t)\}$, allowing the noise of any comb measurement to
be traced unambiguously to the quantum fluctuations of individual comb lines.

\subsection{Quantum noise in optical frequency division}

\begin{figure*}[t]
    \centering
    \includegraphics[width=0.9\textwidth]{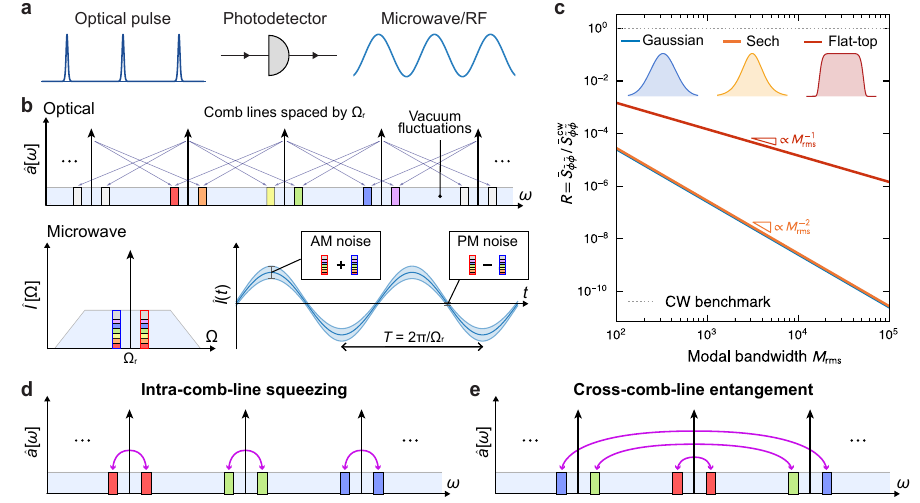}
    \caption{\textbf{Quantum noise in optical frequency division and its suppression.}
    (a)~Principle of OFD: a mode-locked optical pulse train is directly photodetected to
    generate a microwave signal at the repetition frequency $\Omega_r$.
    (b)~Spectral mechanism of quantum noise transduction.
    Top: the optical frequency comb carries vacuum fluctuations (colored boxes) at each
    comb line, spaced by $\Omega_r$; diagonal lines connecting adjacent pairs indicate the
    beating that produces the microwave tone.
    Bottom left: the resulting microwave spectrum at $\Omega_r$, with contributions from
    multiple adjacent-pair beatnotes shown in matching colors.
    Bottom right: decomposition into amplitude-modulation (AM) and phase-modulation (PM)
    noise; AM noise arises from the in-phase sum of adjacent-line vacuum contributions,
    while PM noise arises from their out-of-phase difference, as expressed in
    Eqs.~\eqref{eq:OFD_amp}--\eqref{eq:OFD_phase}.
    (c)~Normalized phase-noise suppression ratio
    $R = \bar{S}_{\tilde{\phi}\tilde{\phi}}/\bar{S}_{\tilde{\phi}\tilde{\phi}}^{\mathrm{cw}}$
    as a function of RMS modal bandwidth $M_{\mathrm{rms}}$ for three archetypal spectral
    envelopes (inset): Gaussian, sech (soliton), and flat-top.
    Dashed line marks the CW benchmark $R = 1$; annotations indicate the
    $M_{\mathrm{rms}}^{-1}$ and $M_{\mathrm{rms}}^{-2}$ asymptotic scalings.
    (d)~Intra-comb-line squeezing for quantum-enhanced OFD.
    Sideband pairs $(\omega_n+\Omega,\,\omega_n-\Omega)$ around each comb line
    are two-mode squeezed (pink arcs) over $|\Omega|<\Omega_r/2$, producing
    phase-quadrature squeezing of the comb-line-resolved field
    $\hat{a}_n[\Omega]$.
    (e)~Cross-comb-line EPR entanglement.
    Sideband pairs $(\omega_0+\Omega',\,\omega_0-\Omega')$ symmetric about the
    global carrier $\omega_0$ are two-mode squeezed (pink arcs). In the comb-line-resolved frame, this correlates
    $\hat{a}_{+n}[\Omega]$ and $\hat{a}_{-n}[-\Omega]$, suppressing the antisymmetric EPR phase quadrature $\hat{P}_n^-$.}
    \label{fig:OFD}
\end{figure*}

Optical frequency division generates ultralow-noise RF/microwave signals
by photodetecting a phase-locked frequency comb [Fig.~\ref{fig:OFD}(a)]. 
From the comb-line-resolved quantum decomposition, one may intuit that
the comb lines act as local oscillators in the photodetection process, heterodyning 
the quantum fluctuations down to the RF/microwave domain, resulting in a quantum noise floor that limits the stability of the generated RF/microwave signal.

To isolate this floor, we substitute the comb field decomposition,
Eq.~\eqref{eq:comb_decomp}, into the photocurrent operator
$\hat{I}(t) = q_e\,\hat{a}^\dagger(t)\hat{a}(t)$ and bandpass-filter
around $\Omega_r$.
Because each $\delta\hat{a}_n(t)$ is spectrally confined to
$|\Omega| \leq \Omega_r/2$, only adjacent-pair cross-terms
($|n-n'|=1$) contribute within the detection window; beats between
non-adjacent lines occur at harmonics $|n-n'|\Omega_r$ and are
rejected by the filter [Fig.~\ref{fig:OFD}(b)].
Retaining these dominant terms:
\begin{multline}
    \hat{I}(t) \approx 2q_e \sum_n \bigl[
    \alpha_{n-1}^* \alpha_n
    + \alpha_{n-1}^*\, \delta\hat{a}_n(t)
    + \alpha_{n+1}\, \delta\hat{a}_n^\dagger(t)
    \bigr] \\
    \times e^{-i\Omega_r t} + \text{H.c.},
    \label{eq:OFD_photocurrent}
\end{multline}
where $\text{H.c.}$ denotes Hermitian conjugate.
Linearizing around the coherent beatnote amplitude
$|S| = |\sum_n \alpha_{n-1}^*\alpha_n|$ and decomposing the quantum
fluctuations into amplitude and phase quadratures,
$\delta\hat{q}_n = (\delta\hat{a}_n + \delta\hat{a}_n^\dagger)/\sqrt{2}$
and $\delta\hat{p}_n = (\delta\hat{a}_n - \delta\hat{a}_n^\dagger)/i\sqrt{2}$,
the photocurrent-linearized estimators of the microwave amplitude and phase fluctuations
for an in-phase comb ($\alpha_n \in \mathbb{R}$) are (see SI; throughout, a tilde denotes
such a photocurrent-derived linear estimator):
\begin{align}
    \delta\tilde{A}(t) &= \frac{1}{\sqrt{2}|S|} \sum_{n}
    \left( \alpha_{n-1} + \alpha_{n+1} \right) \delta\hat{q}_n(t),
    \label{eq:OFD_amp}\\
    \delta\tilde{\phi}(t) &= \frac{1}{\sqrt{2}|S|} \sum_{n}
    \left( \alpha_{n-1} - \alpha_{n+1} \right) \delta\hat{p}_n(t).
    \label{eq:OFD_phase}
\end{align}
Eqs.~\eqref{eq:OFD_amp}--\eqref{eq:OFD_phase} also admit a
frequency-domain interpretation illustrated in Fig.~\ref{fig:OFD}(b):
each $\delta\hat{a}_n[\Omega]$ is heterodyned to two sidebands of
the microwave carrier, with $\delta\hat{a}_n[\Omega]$ (weighted by $\alpha_{n-1}$)
and $\delta\hat{a}_n[\Omega]^\dagger$ (weighted by $\alpha_{n+1}$) forming the
upper and lower sidebands, respectively.
Amplitude noise corresponds to their in-phase combination
$\delta\hat{a}_n+\delta\hat{a}_n^\dagger \propto \delta\hat{q}_n$
with weight $(\alpha_{n-1}+\alpha_{n+1})$, while phase noise corresponds
to their out-of-phase combination
$\delta\hat{a}_n-\delta\hat{a}_n^\dagger \propto \delta\hat{p}_n$
with weight $(\alpha_{n-1}-\alpha_{n+1})$.

\subsubsection{Standard Quantum Limit and spectral shape dependence}

With the comb-line-resolved field $\dop{a}_n[\Omega]$ in the vacuum state for every $n$ and $\Omega$, the (symmetrized) power spectral densities (PSDs) of the quadrature fluctuations are
$\bar{S}_{q_n q_m}[\Omega] = \bar{S}_{p_n p_m}[\Omega] = (1/2)\delta_{nm}$. 
Thus, \cref{eq:OFD_phase} yields a phase noise power spectral density:
\begin{equation}
    \bar{S}_{\tilde{\phi}\tilde{\phi}}
    = \frac{1}{4|S|^2}
      \sum_{n}
      \left(\alpha_{n-1} - \alpha_{n+1}\right)^2,
    \label{eq:OFD_SQL}
\end{equation}
whose frequency-independent character reflects the spectrally uncorrelated character 
of the vacuum across distinct comb lines. The $1/\abs{S}^2$ scaling is the familiar standard 
quantum limit (SQL) for phase estimation with a coherent field of amplitude $|S|$.

Note however that a critical role is played by the factor $(\alpha_{n-1} - \alpha_{n+1})$, 
which is the discrete spectral derivative of the envelope evaluated at line $n$. 
A comb line at a locally flat region of the spectrum
has vanishing derivative and does not contribute to the microwave
phase noise, regardless of its optical power.
Equivalently, a comb with smooth, gradually decaying envelope distributes phase noise sensitivity broadly across all comb lines, 
while those with flat-top spectra concentrate it at the edge lines.

We illustrate this shape dependence by
comparing \cref{eq:OFD_SQL} against the SQL for a microwave tone generated by
heterodyning two CW lasers at the same total photon flux $\abs{S}^2$, for which
$\bar{S}_{\tilde{\phi}\tilde{\phi}}^\mathrm{cw} = 1/4\abs{S}^2$.
When the normalized suppression ratio $R = \bar{S}_{\tilde{\phi}\tilde{\phi}}/\bar{S}_{\tilde{\phi}\tilde{\phi}}^\mathrm{cw}$ is below unity, the comb-based OFD outperforms the CW benchmark.
We also introduce a root mean square (RMS) modal bandwidth, 
\begin{equation}
    M_\mathrm{rms} = \sqrt{\frac{\sum_n n^2\alpha_n^2}{\sum_n \alpha_n^2}}.
    \label{eq:Mrms}
\end{equation}
which weights each line by its squared distance from the carrier and
provides a shape-independent measure of effective spectral reach.
Figure~\ref{fig:OFD}(c) shows $R$ as a function of $M_\mathrm{rms}$
for three archetypal spectral envelopes — Gaussian, sech (soliton), and
flat-top — with insets showing the corresponding pulse shapes.
All three profiles have $R \ll 1$ for any $M_\mathrm{rms} > 1$,
confirming that any spectral broadening beyond two modes is
metrologically beneficial.
The degree of suppression depends critically on how the
envelope tapers toward zero.
The Gaussian and sech envelopes both yield quadratic suppression,
$R \propto M_\mathrm{rms}^{-2}$: their smooth, everywhere-nonzero
spectral derivatives distribute the phase noise sensitivity across all
comb lines, allowing the full optical bandwidth to contribute to
microwave timing stability.
The flat-top envelope achieves only linear suppression,
$R \propto M_\mathrm{rms}^{-1}$.
For a flat-top spectrum, $(\alpha_{n-1}-\alpha_{n+1})$ vanishes for all lines except those
at the edges of the spectrum, so that phase noise arises from the lines at the edges alone
regardless of the optical bandwidth.
The pulse-shape dependence of photodetected comb noise was also identified semiclassically in~\cite{Quinlan2013JOSAB}; here it emerges from the spectral derivative of the comb envelope as a modal-bandwidth-dependent quantum limit.

In SI Sec.~\ref{sec:OFD_classical}, we also use this formalism to rigorously derive the 
classical suppression of the repetition rate phase noise relative to the optical reference $\omega_0$,
showing that the factor $(\Omega_r/\omega_0)^2$ emerges from the spectral structure of the comb and
is independent of the spectral envelope, as long as all comb lines track a common optical reference.

\subsubsection{Quantum-enhanced optical frequency division}

The SQL of \cref{eq:OFD_SQL} can be surpassed by engineering the
quantum state of the comb-line-resolved fields.
Because $\delta\tilde{\phi}$ is a weighted linear combination of the
independent phase quadratures $\{\delta\hat{p}_n\}$, any state that
reduces the variance of this particular linear combination will yield
sub-SQL performance.

The most direct approach is to prepare the comb-line-resolved field $\hat{a}_n[\Omega]$ around each comb line in a phase-quadrature squeezed state with squeezing $1/G_n < 1$, so that the vacuum noise is reduced to $\bar{S}_{p_n p_n} = 1/(2G_n)$ [Fig.~\ref{fig:OFD}(d)].
The resulting phase noise
\begin{equation}
    \bar{S}_{\tilde{\phi}\tilde{\phi}}^{\mathrm{sq}}
    = \frac{1}{4|S|^2}
      \sum_{n}
      \frac{\left(\alpha_{n-1} - \alpha_{n+1}\right)^2}{G_n},
    \label{eq:OFD_sq}
\end{equation}
shows that modes with larger spectral derivatives benefit most
from strong squeezing, while modes on locally flat regions of the
envelope are insensitive to the squeezing level.

A more resource-efficient route exploits the spectral symmetry of the
comb [Fig.~\ref{fig:OFD}(e)].
For a symmetric envelope, $\alpha_n = \alpha_{-n}$, the weights in
Eq.~\eqref{eq:OFD_phase} satisfy
$(\alpha_{n-1} - \alpha_{n+1}) = -(\alpha_{-n-1} - \alpha_{-n+1})$,
so the microwave phase fluctuation depends not on the individual
phase quadratures $\delta\hat{p}_{\pm n}$ but exclusively on their
antisymmetric EPR combination \cite{BrauKimble98},
$\hat{P}_n^- = (\delta\hat{p}_n - \delta\hat{p}_{-n})/\sqrt{2}$.
The symmetric combination $\hat{P}_n^+$ is entirely decoupled from the phase measurement; its quantum statistics may take any form without penalty to the OFD measurement.
Broadband two-mode squeezing of all sideband pairs
$(\omega_{+n}+\Omega,\,\omega_{-n}-\Omega)$ with uniform gain $G_n$
over $|\Omega|<\Omega_r/2$---equivalently, cross-comb-line entanglement of
the comb-line-resolved fields $\hat{a}_{+n}[\Omega]$ and
$\hat{a}_{-n}[-\Omega]$---reduces the antisymmetric EPR quadrature noise to $\bar{S}_{P_n^- P_n^-} = 1/2G_n$,
yielding the phase noise
\begin{equation}
    \bar{S}_{\tilde{\phi}\tilde{\phi}}^{\mathrm{ent}}
    = \frac{1}{2|S|^2}
      \sum_{n \geq 1}
      \frac{\left(\alpha_{n-1} - \alpha_{n+1}\right)^2}{G_n},
    \label{eq:OFD_ent}
\end{equation}
which matches \cref{eq:OFD_sq} under the symmetric-envelope assumption.

\subsection{Quantum noise in dual-comb spectroscopy}

\begin{figure*}[t]
    \centering
    \includegraphics[width=\textwidth]{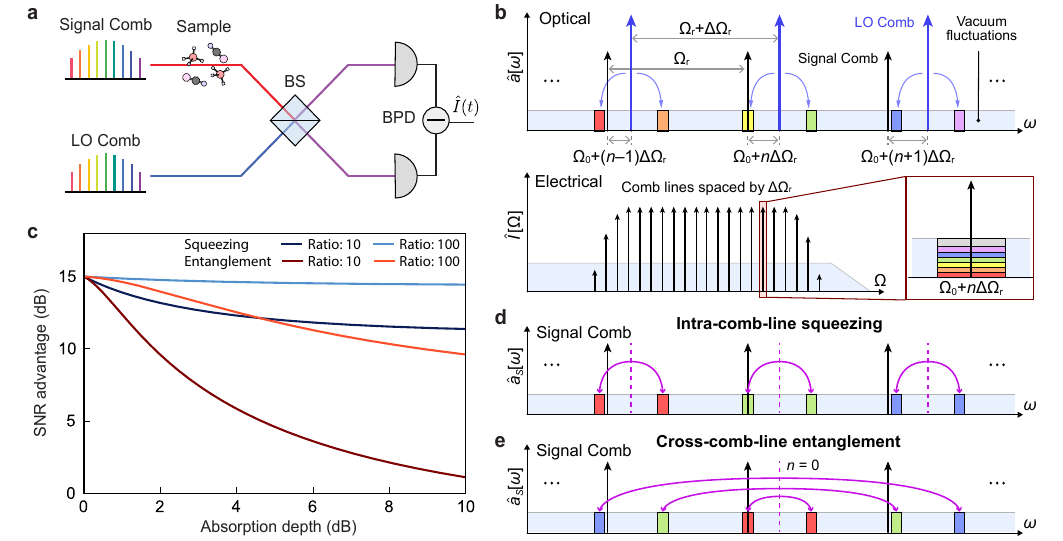}
    \caption{\textbf{Quantum noise in dual-comb spectroscopy and its suppression.}
    (a)~Experimental setup: a signal comb passes through a molecular sample and combines with a local oscillator (LO) comb at a 50:50 beam splitter (BS), and a balanced photodetector (BPD) records the difference photocurrent $\hat{I}(t)$.
    (b)~Frequency-domain vacuum noise folding in DCS. Top (optical): the signal comb (black, line spacing $\Omega_r$) is shown together with the LO comb (blue, spacing $\Omega_r+\Delta\Omega_r$), with colored regions marking the signal-comb vacuum fluctuations at offsets $\Omega_0+n\Delta\Omega_r$ relative to the LO comb lines. Bottom (electrical): beating during photodetection maps each pair to an RF comb tone spaced by $\Delta\Omega_r$, and the inset shows how vacuum contributions from signal comb incoherently add at the same RF frequency $\Omega_0+n\Delta\Omega_r$.
    (c)~SNR advantage $10\log_{10}\mathcal{A}_\mathrm{SNR}$ (dB) of intra-comb-line squeezing (blue) and cross-comb-line EPR entanglement (red) as a function of single-line absorption depth, with a 15 dB squeezing level ($G \approx 31.6$), for ratios of intact-to-absorbed comb lines of $10$ and $100$.
    (d)~Intra-comb-line squeezing.
    Sideband pairs $(\omega_n+\Omega_0+\Omega,\,\omega_n+\Omega_0-\Omega)$ around each nearest LO comb line are two-mode squeezed (pink arcs), producing amplitude-quadrature squeezing of the cross-referred field $\hat{a}_{S,n}^{\rm CR}[\Omega]$.
    (e)~Cross-comb-line EPR entanglement.
    Sideband pairs $(\omega_0+\Omega_0+\Omega',\,\omega_0+\Omega_0-\Omega')$ around the central LO comb line are two-mode squeezed (pink arcs), correlating cross-referred signal fields $\hat{a}_{S,\pm m}^{\rm CR}$ and suppressing the symmetric EPR amplitude quadrature $\hat{Q}^+_{S,m}$.}
    \label{fig:DCS}
\end{figure*}

We now turn to dual-comb spectroscopy (DCS), wherein broadband high-resolution spectroscopy is performed by beating two frequency combs against each other after one of them has interacted with a sample.
Quantum-enhanced DCS has recently been analyzed within discrete-mode formulations: to overcome the rotating measurement quadrature inherent to multi-heterodyne detection, cross-referred squeezing was first proposed in~\cite{Shi23}, cross-comb-line entanglement in~\cite{Hariri25}, and a unified theory of these protocols in~\cite{Zhuang25}.
Our continuous-mode treatment below provides an independent, first-principles validation of these strategies and traces the spoiling of straightforward squeezing to a time-domain cyclostationary noise penalty---recently observed experimentally~\cite{HermanPhaseDep25}---yielding a rigorous mechanistic account.
Importantly, the formalism is more naturally suited to capture the effects of classical technical noise and the role of integration time, both of which are practically relevant.

In DCS, one comb with repetition rate $\Omega_r$ serves as the signal comb, which 
propagates through the sample, while the other comb with repetition
rate $\Omega_r + \Delta\Omega_r$ serves as a local oscillator (LO)
[Fig.~\ref{fig:DCS}(a)]. Propagation through the sample maps each per-line field $\hat{a}_{S,n}$ as
$\hat{a}_{S,n} \mapsto \sqrt{\kappa_n}e^{i\theta_n}\hat{a}_{S,n}
 + \sqrt{1-\kappa_n}\hat{v}_n$
(see SI Sec.~\ref{sec:dualcomb}), where $\kappa_n\in[0,1]$ is the
intensity transmittance, $\theta_n$ is the phase delay, and
$\hat{v}_n$ is an uncorrelated noise field, ideally in the vacuum
state; the $\sqrt{1-\kappa_n}\,\hat{v}_n$ term enforces the canonical
commutation relations. After this sample interaction, the transmitted
signal and LO fields are recombined on a 50:50 beam splitter and
measured by a balanced photodetector, which records the
cross-correlation photocurrent
\begin{equation}
    \hat{I}(t) = q_e\bigl[
        \hat{a}_S^\dagger(t)\hat{a}_L(t)
        + \hat{a}_L^\dagger(t)\hat{a}_S(t)
    \bigr],
    \label{eq:DCS_balanced}
\end{equation}
which we analyze in the strong-LO limit
$|\alpha_{L,n}|\gg|\alpha_{S,n}|$, where the bright LO acts as a
classical reference that amplifies the weak signal.

Each pair of signal and LO comb lines indexed by $n$ generates one RF
beat note in the photocurrent at frequency $\Omega_n=\Omega_0+n\,\Delta\Omega_r$, where
$\Omega_0$ is the carrier offset between the two combs; its amplitude
and phase encode the sample transmittance and delay at the optical frequency
$\omega_n=\omega_0+n\Omega_r$~\cite{Coddington2016}.
This RF comb is densely packed: the tone spacing $\Delta\Omega_r$ is many
orders of magnitude smaller than the per-line optical bandwidth
$\Omega_r$.
Consequently, the noise at any RF detection frequency $\Omega$ is due to
each LO line $n$ beating independently with the signal vacuum at optical offset $\Omega$
from itself, folding an independent vacuum contribution from every
signal comb line into the same RF frequency [Fig.~\ref{fig:DCS}(b)].

To make the quadrature structure of this folding explicit, we momentarily
set the sample aside ($\kappa_n=1$, $\theta_n=0$) and write the
strong-LO photocurrent fluctuation in the natural ``self-referred''
frame, where each signal line's $\delta\hat{q}_{S,n}$,
$\delta\hat{p}_{S,n}$ are defined relative to its own optical carrier
$\omega_n$:
\begin{equation}
    \delta\hat{I}(t) \approx
    \sqrt{2}\,q_e\sum_n \alpha_{L,n}
    \bigl[
        \delta\hat{q}_{S,n}(t)\cos\Omega_n t
        -\delta\hat{p}_{S,n}(t)\sin\Omega_n t
    \bigr].
    \label{eq:DCS_SR}
\end{equation}
Clearly, the photocurrent is not determined by a fixed quadrature of the signal comb;
rather, the LO carrier defines the detected quadrature which rotates at the beat frequency $\Omega_n$ relative to the signal's own carrier.
This has no consequence when the signal comb is in the vacuum state: in this case,
the quadrature variances are isotropic ($\bar{S}_{q_{S,n}q_{S,n}}=\bar{S}_{p_{S,n}p_{S,n}}=1/2$)
resulting in a photocurrent that is stationary with a white spectrum (see SI).
However, when the signal comb is squeezed in this self-referred frame, the isotropy is
broken ($\bar{S}_{q_{S,n}q_{S,n}}=1/(2G), \bar{S}_{p_{S,n}p_{S,n}}=G/2$ for amplitude squeezing),
and the photocurrent alternately samples the squeezed and anti-squeezed quadratures, 
such that its variance oscillates between the squeezed level $1/G$ and the anti-squeezed level $G$ at twice the beat frequency $\Omega_n$ (see SI).
That is, the photocurrent is cyclostationary, and its time-averaged power spectrum 
is increased by the factor $(G+1/G)/2 \geq 1$ relative to the vacuum case, eliminating any quantum advantage.
The failure of self-referred squeezing under heterodyne was recognized in frequency-domain in~\cite{Shi23}; here it acquires a time-domain origin as cyclostationary noise, recently observed in Kerr-squeezed dual-comb interferograms~\cite{HermanPhaseDep25}.

Effective quantum enhancement therefore requires aligning the squeezing
axis with the quadrature the detector actually reads out---that is,
with the LO-defined measurement axis.
The ``cross-referred'' frame~\cite{Shi23} achieves precisely this by
proposing a signal frame,
$\delta\hat{a}_{S,n}^{\mathrm{CR}}(t)\equiv \delta\hat{a}_{S,n}(t)\,e^{i\Omega_n t}$,
that is co-rotating with the LO carrier at $\omega_n + \Omega_n$ (rather than the self-referred
frame that is co-rotating with the signal carrier at $\omega_n$).
The signal quadrature in this frame
$\delta\hat{q}_{S,n}^{\mathrm{CR}}(t) =
\delta\hat{q}_{S,n}(t)\cos\Omega_n t
-\delta\hat{p}_{S,n}(t)\sin\Omega_n t$
is tracked by the LO, and the resulting photocurrent 
reduces to a single stationary observable
[Fig.~\ref{fig:DCS}(d)--(e)]:
\begin{equation}
    \delta\hat{I}(t) \approx \sqrt{2}\,q_e \sum_n
        \alpha_{L,n}\,\delta\hat{q}_{S,n}^{\mathrm{CR}}(t).
    \label{eq:DCS_photocurrent_LO}
\end{equation}
The phase quadrature is now fully decoupled from the photocurrent, and
suppressing the variance of $\delta\hat{q}_{S,n}^{\mathrm{CR}}$ directly
suppresses every RF beat note's noise floor without incurring any
cyclostationary noise or anti-squeezing penalty.

\subsubsection{Standard quantum limit}

With the signal field in the vacuum state for every $n$ and $\Omega$, $\bar{S}_{q_{n}q_{n}} = 1/2$,
Eq.~\eqref{eq:DCS_photocurrent_LO} yields a white photocurrent PSD:
\begin{equation}
    \bar{S}_{II}^{\mathrm{SQL}}[\Omega]
    = q_e^2 \sum_n \alpha_{L,n}^2.
    \label{eq:DCS_SQL}
\end{equation}
Substituting $\bar{S}_{II}^{\mathrm{SQL}}$ into the SNR for the
transmittance estimator $\hat{\kappa}_m$ (see SI Sec.~\ref{sec:SNR_kappa})
gives
\begin{equation}
    \mathrm{SNR}_{\kappa_m}^{\mathrm{SQL}}
    = \frac{\kappa_m\,\alpha_{S,m}^2\,\alpha_{L,m}^2\,T}
           {2\displaystyle\sum_n \alpha_{L,n}^2}.
    \label{eq:DCS_SNR_SQL}
\end{equation}
The denominator accumulates vacuum noise from all $2N+1$ LO comb lines, not just line $m$: multi-heterodyne folding forces every absorption measurement to contend with the full comb's shot noise floor, making quantum noise reduction the central challenge in DCS precision.
The SNR grows linearly with integration time $T$, reflecting the $1/\sqrt{T}$ averaging of white photocurrent noise.

\subsubsection{Quantum-enhanced dual-comb spectroscopy}

The structure of Eq.~\eqref{eq:DCS_photocurrent_LO} identifies the
amplitude quadrature variance $\bar{S}_{q_{S,n}^{\mathrm{CR}} q_{S,n}^{\mathrm{CR}}}$ as the relevant
quantum resource in suppressing the photocurrent noise. 
(Squeezing the LO comb is not beneficial because the LO vacuum fluctuations do not contribute to the photocurrent noise in the strong-LO limit.)
Squeezing the signal's cross-referred amplitude quadrature with per-line
gain $G_{S,n}$ gives
[Fig.~\ref{fig:DCS}(d)]:
\begin{equation}
    \bar{S}_{II}^{\mathrm{sq}}[\Omega]
    = q_e^2 \sum_n \alpha_{L,n}^2 \Bigl(
        \frac{\kappa_n \cos^2\theta_n}{G_{S,n}}
        + G_{S,n}\,\kappa_n\sin^2\theta_n
        + (1 - \kappa_n)
    \Bigr).
    \label{eq:DCS_CR}
\end{equation}
The first term is the squeezed contribution, an improvement by the factor $G_{S,n}$ relative to the SQL; the third term is the vacuum contribution from the sample loss.
The second term, due to anti-squeezing that is rotated into the measurement quadrature by
the sample phase delay $\theta_n$, can degrade the advantage due to squeezing (analogous
to the degradation in single-mode squeezing due to phase noise \cite{DwyeSigg13}). 
However, quantum enhancement persists when the phase delay $\theta_n$ is small---a condition satisfied in weakly absorbing molecular samples where the phase shift is negligible~\cite{herman25} or perhaps by actively pre-compensating the phase using a waveshaper~\cite{Jiang07}.

An alternative approach exploits quantum correlations between symmetric
line pairs [Fig.~\ref{fig:DCS}(e)].
For a symmetric comb envelope ($\alpha_{J,n} = \alpha_{J,-n}$), the
photocurrent Eq.~\eqref{eq:DCS_photocurrent_LO}
is exclusively sensitive to the symmetric EPR amplitude quadrature of the
signal comb,
$\hat{Q}_{S,m}^+ = (\delta\hat{q}_{S,+m}^{\t{CR}} + \delta\hat{q}_{S,-m}^{\t{CR}})/\sqrt{2}$:
\begin{equation}
    \delta\hat{I}(t)
    \approx \sqrt{2}\,q_e\,\alpha_{L,0}\,\delta\hat{q}_{S,0}^{\mathrm{CR}}(t)
    + 2q_e \sum_{m=1}^{N}
        \alpha_{L,m}\,\hat{Q}_{S,m}^+(t),
    \label{eq:DCS_EPR}
\end{equation}
with the conjugate quadrature $\hat{Q}^-_{S,m}=(\delta\hat{q}_{S,+m}^{\t{CR}} - \delta\hat{q}_{S,-m}^{\t{CR}})/\sqrt{2}$ decoupled.
Two-mode squeezing of sideband pairs ($\dop{a}_{S,+m}^{\t{CR}}[\Omega]$, $\dop{a}_{S,-m}^{\t{CR}}[-\Omega]$) can exclusively 
suppress the noise from $\hat{Q}^+$.
The scheme is, however, fragile to spectral asymmetry: any asymmetric
absorption between the $+m$ and $-m$ lines 
couples the anti-squeezed $\hat{Q}^-$ quadrature into the photocurrent,
whose contribution grows with $G_{S,m}$ and the degree of
absorption asymmetry (see SI Eq.~\eqref{eq:PSD_EPR_pair}
for the exact photocurrent PSD).

Figure~\ref{fig:DCS}(c) compares the SNR advantage factor
$\mathcal{A}_\mathrm{SNR} \equiv \bar{S}_{II}|_{G=1} / \bar{S}_{II}$
as a function of single-line absorption depth for both strategies
in a localized absorber model (see SI).
For intra-comb-line squeezing, the anti-squeezed quadrature is never admitted to the photocurrent; as absorption increases, only vacuum injected by the sample contributes an additive $(1-\kappa)$ penalty that is independent of $G_{S,n}$.
For cross-comb-line entanglement, absorbing only one of the entangled pair breaks the bipartite symmetry, directly injecting the anti-squeezed quadrature $\hat{Q}^-$ into the photocurrent with a penalty that grows with $G_{S,m}$.
Intra-comb-line squeezing is therefore the robust quantum resource for samples with localized or spectrally asymmetric absorption features.

\section{Discussion}
We have developed a framework for analyzing the quantum noise of frequency comb metrology
that is developed from a first-principles quantization of the continuous comb field using a 
rigorous comb-line resolved decomposition. In this framework, quantum and classical (technical) 
noise enters on equal footing through the 
substitution $\delta\hat{q}_n[\Omega] \to \delta\hat{q}_n[\Omega] 
+ \delta\hat{q}_n^{\mathrm{cl}}[\Omega]$ (and likewise for $\delta\hat{p}_n$), where $\delta\hat{q}_n^{\mathrm{cl}}$ and $\delta\hat{p}_n^{\mathrm{cl}}$  encode the classical amplitude and phase noise on the $n$-th comb line, respectively, thereby unifying their treatment in a seamless manner.

We have applied this formalism to two canonical examples. 
In OFD, it reveals that the phase-noise SQL scales as $M_\mathrm{rms}^{-2}$ for smooth comb envelopes, but only as $M_\mathrm{rms}^{-1}$ for flat-top spectra, and the efficient quantum resources for surpassing these limits are either intra-comb-line phase quadrature squeezing or cross-comb-line entanglement in antisymmetric EPR phase quadratures.
In DCS, it shows that self-referred squeezing incurs a cyclostationary noise penalty that can be circumvented by cross-referred squeezing~\cite{Shi23} or cross-comb-line EPR entanglement in symmetric EPR amplitude quadratures~\cite{Hariri25}.
The formalism and resulting insights are broadly applicable to comb-based measurement modalities, including laser ranging~\cite{Coddington09}, astronomical spectrograph calibration~\cite{Steinmetz08}, and timekeeping.

The experimental realization of quantum-enhanced comb metrology requires platforms capable of generating the requisite quantum states across the full comb bandwidth.
Both $\chi^{(2)}$ and $\chi^{(3)}$ parametric processes provide natural routes to multimode squeezed and entangled states distributed across the comb spectrum~\cite{Roslund14,Fabre20,
Yang21,Nehra22,Jankowski20,herman25,Wang25}.
Microresonator Kerr combs are a particularly compelling match to the cross-comb-line entanglement resource identified here: the four-wave mixing that seeds comb generation automatically entangles symmetric sideband pairs $(\pm n)$~\cite{Chembo16,Yang21}, structurally identical to the states that minimize the OFD and DCS noise floors.
A key open question is which multimode quantum state a given nonlinear platform will actually generate. While recent advances in Gaussian split-step Fourier methods~\cite{Ng23} and efficient quantum process tomography~\cite{Ra25} provide a powerful blueprint for analyzing discrete multimode systems, the path forward lies in generalizing these tools to capture and engineer the continuous offset-frequency correlations that fundamentally govern practical metrology.
Together with the present theory---which maps these continuous correlations directly to a metrological advantage---such developments suggest that the full co-optimization of comb platform, nonlinear interaction, and measurement architecture as a unified quantum system is now within experimental reach.

\bibliography{refs_comb_quantum}


\clearpage
\onecolumngrid
\appendix

\begin{center}
    \textbf{\large Supplementary Information}
\end{center}

\writetoctrue
\tableofcontents

\section{Operator conventions and spectral densities}\label{sec:spectral_densities}

We define the Fourier transform of a time-dependent operator $\hat{A}(t)$ in the Heisenberg picture as
\begin{equation}\label{Fourier}
    \hat{A}[\Omega] = \int_{-\infty}^{+\infty} \hat{A}(t) e^{i\Omega t} \, dt,
\end{equation}
with inverse transform
\begin{equation}
    \hat{A}(t) = \int_{-\infty}^{+\infty} \hat{A}[\Omega]\, e^{-i\Omega t} \frac{d\Omega}{2\pi}.
\end{equation}
The Hermitian conjugate and Fourier transform do not commute: $\hat{A}[\Omega]^\dagger = \hat{A}^\dagger[-\Omega]$.

The unsymmetrized cross-spectrum of fluctuation operators $\delta\hat{A}$, $\delta\hat{B}$ is
\begin{equation}
    S_{AB}[\Omega] = \int_{-\infty}^{+\infty}
    \left\langle \delta\hat{A}^\dagger(t)\,\delta\hat{B}(0) \right\rangle e^{i\Omega t}\,dt.
    \label{eq:CrossSpec}
\end{equation}
For wide-sense stationary operators this reduces to
\begin{equation}
    \langle\delta\hat{A}^\dagger[\Omega]\,\delta\hat{B}[\Omega']\rangle
    = S_{AB}[\Omega]\cdot 2\pi\,\delta[\Omega+\Omega'].
\end{equation}
The symmetrized cross-spectral density,
\begin{equation}
    \begin{split}
    \bar{S}_{AB}[\Omega] \equiv \int_{-\infty}^{+\infty}
    \left\langle \tfrac{1}{2}\bigl\{\delta\hat{A}^\dagger(t),\,\delta\hat{B}(0)\bigr\} \right\rangle
    e^{i\Omega t}\,dt\\
    = \tfrac{1}{2}\bigl(S_{AB}[\Omega] + S_{B^\dagger A^\dagger}[-\Omega]\bigr),
    \end{split}
\end{equation}
is real and even for Hermitian observables: $\bar{S}_{XX}[\Omega]=\bar{S}_{XX}[-\Omega]$.
The single-sided PSD $\bar{S}_X[\Omega]:=2\bar{S}_{XX}[\Omega]$ for $\Omega\geq 0$ satisfies
\begin{equation}
    \mathrm{Var}[\delta\hat{X}] = \int_0^\infty \bar{S}_X[\Omega]\frac{d\Omega}{2\pi}.
\end{equation}

\section{Quantum model of optical frequency combs}
\subsection{Quantization of frequency combs}
The quantum theory of propagating electromagnetic fields is most naturally formulated in a basis of continuous spatiotemporal modes, which avoids artifacts associated with cavity quantization. We review this formalism by quantizing the electromagnetic field in a finite volume and extending this description to an infinite one-dimensional space, which is characteristic of free-space optical systems.

The quantization procedure begins with the vector potential operator $\hat{\bm{A}}(\bm{r}, t)$ in a quantization volume $V=L^3$. In the Coulomb gauge, the operator is expanded over a complete set of discrete plane-wave modes with wave vectors $\bm{k}_i$. By taking the continuum limit ($L \to \infty$), the summation over discrete modes transitions to an integral over a continuous $\bm{k}$-space, according to the rule $\frac{1}{V}\sum_{\bm{k}} \to \int \frac{d^3k}{(2\pi)^3}$. The vector potential operator for the free field is then given by \cite{Cohen97}:
\begin{equation} \label{eq:A_3D}
\hat{\bm{A}}(\bm{r},t)= \int \frac{d^3k}{(2\pi)^3} \sum_{\lambda=1,2} \bm{\epsilon}(\bm{k},\lambda) \sqrt{\frac{\hbar}{2\epsilon_0\omega_{\bm{k}}}}
\left[\hat{a}_\lambda[\bm{k}] e^{i(\bm{k}\cdot\bm{r} - \omega_{\bm{k}}t)}+\hat{a}_\lambda[\bm{k}]^\dagger e^{-i(\bm{k}\cdot\bm{r} - \omega_{\bm{k}}t)}\right],
\end{equation} 
where $\omega_{\bm{k}}=c|\bm{k}|$, $\bm{\epsilon}(\bm{k},\lambda)$ is the polarization vector, $\hat{a}_\lambda[\bm{k}]$ is the annihilation operator for a mode with wave vector $\bm{k}$ and polarization $\lambda$, and $\hat{a}_\lambda[\bm{k}]^\dagger$ is the corresponding creation operator.

For many applications, it is sufficient to consider a one-dimensional model where the field propagates along the $z$-axis within a finite cross-sectional area $\mathcal{A}$.
This simplification is equivalent to considering only modes with transverse wave-vector components $k_x=k_y=0$. The integral over $d^3k$ thus reduces to a one-dimensional integral over $k_z \equiv k$, with the transformation $\int d^3k \to \frac{(2\pi)^2}{\mathcal{A}}\int dk$. Assuming a single, linear polarization, the vector potential becomes:
\begin{equation} \label{eq:A_1D_k}
\hat{A}(z,t)=\int_{-\infty}^{\infty} \frac{dk}{2\pi} \sqrt{\frac{\hbar}{2\mathcal{A}\epsilon_0  c|k|}}
\left[\hat{a}[k] e^{i(kz - c|k|t)}+\text{H.c.}\right],
\end{equation}
where $\text{H.c.}$ denotes Hermitian conjugate.
The canonical commutation relation for these one-dimensional k-space operators is 
\begin{equation}\label{eq:commu_k}
    [\hat{a}[k], \hat{a}[k']^\dagger] = 2\pi\,\delta[k-k'].
\end{equation}
This relation defines a continuum of bosonic modes, allowing the electromagnetic field to be treated as a collection of independent quantum harmonic oscillators, with photons being the quantized excitations.

We further restrict our analysis to fields propagating only in the positive $z$-direction, for which $k>0$. It is convenient to transform the description from wavenumber space to frequency space using the linear dispersion relation $\omega=ck$. 
This change of basis requires a careful redefinition of the field operators. 
In this quantum description, the operator $\hat{a}^\dagger[k]\hat{a}[k]$ is interpreted as the photon number density in wavenumber space.
To preserve the total photon number under the change of basis, the equality $\hat{a}^\dagger[\omega]\hat{a}[\omega]d\omega = \hat{a}^\dagger[k]\hat{a}[k]dk$ must hold.
This implies the operator transformation $\hat{a}[\omega] = \hat{a}[k]/\sqrt{c}$.
Therefore, the vector potential can also be written as \cite{Blow90}:
\begin{equation} \label{eq:A_1D_omega}
\hat{A}(z,t) = \int_0^\infty \frac{d\omega}{2\pi}\sqrt{\frac{\hbar}{2\mathcal{A}\epsilon_0  c \omega}} \left[\hat{a}[\omega] e^{-i\omega(t-z/c)}+\text{H.c.}\right].
\end{equation}
The electric field operator is then derived from this expression via $\hat{\bm{E}} = -\partial_t \hat{\bm{A}}$:
\begin{equation} \label{eq:E_1D_omega}
\hat{E}(z,t) = \int_0^\infty \frac{d\omega}{2\pi} \sqrt{\frac{\hbar\omega}{2\mathcal{A}\epsilon_0 c}} \left[i\hat{a}[\omega] e^{-i\omega(t-z/c)}+\text{H.c.}\right].
\end{equation}
Since the field of a wave propagating along the z-axis is simply a time-delayed version of its value at the origin, its behavior can be fully described by a single function of time, $\hat{A}(t)\eqdef \hat{A}(0,t)$ and $\hat{E}(t)\eqdef \hat{E}(0,t)$ as
\begin{align}
    \hat{A}(t) &= \int_0^\infty \frac{d\omega}{2\pi}\sqrt{\frac{\hbar}{2\mathcal{A}\epsilon_0  c \omega}} \left[\hat{a}[\omega] e^{-i\omega t}+\text{H.c.}\right].\\
     \hat{E}(t) &= \int_0^\infty \frac{d\omega}{2\pi} \sqrt{\frac{\hbar\omega}{2\mathcal{A}\epsilon_0 c}} \left[i\hat{a}[\omega] e^{-i\omega t}+\text{H.c.}\right].\label{eq:E(t)_q}
\end{align}
Using the commutation relation in the wavenumber space shown in \cref{eq:commu_k}, we derive the commutation relation in the frequency domain:
\begin{equation} \label{eq:commutation_omega}
\begin{split}
[\hat{a}[\omega], \hat{a}[\omega']^\dagger] = 2\pi\, \delta[\omega-\omega'].
\end{split}
\end{equation}

We now consider a quantum model for an optical frequency comb. Our focus is on the quantum limits of metrological applications where the measurement integration time is much longer than the pulse-to-pulse period. This long-time observation window justifies modeling the comb not as a single pulse, but as an infinite, periodic train of pulses; that is, a coherent polychromatic field in the frequency domain operating in a steady state.

Classically, the electric field of such a pulse train can be described in the time domain as a periodic envelope function, $E_{\text{env}}(t)$, modulating a carrier wave at frequency $\omega_0$:
\begin{equation}
E^{(+)}(t) = E_{\text{env}}(t) e^{-i\omega_0 t},
\end{equation}
where we decompose $E=E^{(+)}+E^{(-)}$ with $E^{(+)}=E^{(-)\dagger}$.
The envelope $E_{\text{env}}(t)$ is periodic with a repetition period $T_r = 2\pi/\Omega_r$, where $\Omega_r$ is the (angular) repetition rate.
As a periodic function, it can be expanded as a Fourier series:
\begin{equation}
E_{\text{env}}(t) = \sum_{n=-\infty}^{\infty} E_n e^{-in\Omega_r t}
\end{equation}
where the complex coefficients $E_n$ are the spectral shape of the pulse envelope, defined as $E_m=(1/T_r)\int^{T_r/2}_{-T_r/2}E_{\text{env}}(t)e^{im\Omega_r t}$.
Substituting the series expansion back into the expression for the field and taking the Fourier transform gives a series of discrete, equally-spaced spectral lines—a frequency comb:
\begin{equation}
E^{(+)}[\omega] = \sum_{n=-\infty}^{\infty} E_n\, 2\pi\, \delta[\omega - \omega_n],
\end{equation}
where $\omega_n = \omega_0 + n\Omega_r$.

To construct the quantum model, we require that the expectation value of the electric field operator, $\langle \hat{E}(t) \rangle$, reproduces the classical field $E(t)$. From \cref{eq:E(t)_q}, we can relate the classical spectrum $\tilde{E}^{(+)}[\omega]$ to the coherent amplitude $\alpha[\omega] \equiv \langle \hat{a}[\omega] \rangle$, i.e.,
\begin{equation}
    \alpha[\omega] = -i\sqrt{\frac{2\mathcal{A}\epsilon_0 c}{\hbar \omega}}\sum_{n=-\infty}^{\infty} E_n\, 2\pi\, \delta[\omega - \omega_n].
\end{equation}
The full quantum operator $\hat{a}[\omega]$ is then expressed as the sum of this classical amplitude and a quantum fluctuation operator:
\begin{equation}
\hat{a}[\omega] = \alpha[\omega] + \delta\hat{a}[\omega],
\end{equation}
where $\delta\hat{a}[\omega]$ is a bosonic operator, which satisfies the canonical commutation relation (\cref{eq:commutation_omega})
\begin{equation}\label{eq:comm_delta}
[\delta\hat{a}[\omega], \delta\hat{a}[\omega']^\dagger] = 2\pi\,\delta[\omega-\omega']
\end{equation}
For simplicity, we use the form of  $\alpha[\omega]=\sum_{n=-\infty}^{\infty}\alpha_n\,2\pi\,\delta[\omega-\omega_n]$ in the following sections.

\subsection{Broadband Photodetection}\label{sec:broadband_SI}
We model the photodetection of broadband light using the framework of Yurke \cite{yurke85}, in which the photodiode is described by a photoemission--rate operator $\hat{W}(t)$ and the measured current is the filtered version of $e\,\hat{W}(t)$. In the frequency domain we write
\begin{equation}
\hat{I}[\Omega]=q_e\,H[\Omega]\,\hat{W}[\Omega],\quad\hat{I}(t)=\int_0^\infty \frac{d\Omega}{2\pi}\,\big[\hat{I}[\Omega]e^{-i\Omega t}+\text{H.c.}\big],
\end{equation}
where $q_e$ is the charge of an electron and $H[\Omega]$ is the transfer function of the electronics. Throughout, $\Omega$ denotes the photocurrent offset frequency within the electronics passband (RF--microwave range), which is much smaller than optical frequencies.

The wideband photoemission operator $\hat{W}[\Omega]$ is a bilinear functional of optical frequency components separated by $\Omega$:
\begin{equation}
\hat{W}[\Omega]=\int_{0}^{\infty} \frac{d\omega}{2\pi} F[\Omega,\omega]\;\hat{a}[\omega]^\dagger\hat{a}[\omega+\Omega].
\label{eq:W-bilinear-general}
\end{equation}
The kernel $F[\Omega,\omega]$ is determined by the intrinsic sensitivity of the detector. Assuming the experimentally relevant separation of scales, $\Omega \ll \omega$, the detection kernel becomes flat ($F[\Omega,\omega] \approx 1$), such that
\begin{equation}\label{eq:photoemission-simple}
\hat{W}[\Omega]=\int_{0}^{\infty}\! \frac{d\omega}{2\pi}\; \hat{a}[\omega]^\dagger\hat{a}[\omega+\Omega],
\end{equation}
where the lower integration limit is extended from a cutoff frequency $\omega_c$ to zero for mathematical convenience.

Next, we derive the time-domain representation of the photocurrent, $\hat{I}(t)$. This involves finding the inverse Fourier transform of the photoemission operator, $\hat{W}(t)$, and applying the convolution theorem.
It is useful to define the time-domain annihilation operator $\hat{a}(t)$ as the inverse Fourier transform of the (positive-frequency) operator $\hat{a}[\omega]$:
\begin{equation}\label{eq:time_annhi}
\hat{a}(t) = \int_0^\infty \frac{d\omega}{2\pi} \hat{a}[\omega] e^{-i\omega t},\quad \hat{a}^\dagger(t) = \int_0^\infty \frac{d\omega}{2\pi} \hat{a}[\omega]^\dagger e^{i\omega t}.
\end{equation}
The inverse Fourier transform for $\hat{W}(t)$ is given as
\begin{equation}
    \hat{W}(t) = \int_{-\infty}^{\infty} \frac{d\Omega}{2\pi} \hat{W}[\Omega] e^{-i\Omega t}.
\end{equation}
Substitute the simplified expression for $\hat{W}[\Omega]$ from \cref{eq:photoemission-simple}:
\begin{align}
   \hat{W}(t) &= \int_{-\infty}^{\infty} \frac{d\Omega}{2\pi} \left( \int_0^\infty \frac{d\omega}{2\pi} \hat{a}[\omega]^\dagger \hat{a}[\omega+\Omega] \right) e^{-i\Omega t}\\
   &=\int_0^\infty \frac{d\omega}{2\pi} \hat{a}[\omega]^\dagger \left( \int_{-\infty}^{\infty} \frac{d\Omega}{2\pi} \hat{a}[\omega+\Omega] e^{-i\Omega t} \right)\\
   &=\left( \int_0^\infty \frac{d\omega}{2\pi} \hat{a}[\omega]^\dagger e^{i\omega t} \right) \hat{a}(t)= \hat{a}^\dagger(t)\hat{a}(t).
\end{align}
The final step in this derivation relies on a change of variables $\omega'=\omega+\Omega$; the resulting integral over $\omega'$ is taken from 0 to $\infty$, which is justified because the annihilation operator $a[\omega']$ for the positive-frequency field is identically zero for $\omega'<0$. 
It is important to recognize that this simple, local form for the photoemission operator is a direct consequence of the flat-detector-response approximation $\Omega\ll\omega$ made earlier.
With the time-domain photoemission operator established, the photocurrent operator is found by convolution with the electronics' impulse response $h(t)$, defined as the inverse Fourier transform of the transfer function $H[\Omega]$. 
That is,
\begin{equation}\label{eq:photocurrent}
\hat{I}(t) = q_e \int_{-\infty}^{\infty} d\tau\, h(t-\tau) \hat{a}^\dagger(\tau)\hat{a}(\tau),
\end{equation}
for an ideal detector.
This shows that the photocurrent is the electronically filtered version of the instantaneous photon flux operator. 

As shown above, the time-domain annihilation operator defined in \cref{eq:time_annhi} provides a simple framework for understanding the dynamics of the photocurrent. In all of the analysis that follows, we compute the photocurrent operator by obtaining $\hat{a}^\dagger(t)\hat{a}(t)$ at the detector and lowpass filtering it with an associated cutoff frequency; for example, the cutoff frequency is $\Omega_r/2$ in dual-comb spectroscopy.

\subsection{Comb-line-resolved field decomposition}
\label{app:comb_line_resolved}

As will be shown in \cref{sec:OFD_SI,sec:dualcomb}, the quantum noise floor
of frequency-comb metrology is determined by the vacuum fluctuation in a
narrow baseband window around each comb line. It is therefore natural to
decompose the continuous-mode operator $\hat{a}[\omega]$ into a discrete
family $\{\hat{a}_n[\Omega]\}$, where $\hat{a}_n[\Omega]$ describes
fluctuations at RF offset $\Omega$ around the $n$-th optical comb line
$\omega_n = \omega_0 + n\Omega_r$. In this section we introduce the most
general linear decomposition of this form, identify the two conditions
under which it constitutes a unitary change of basis, and show that
within the comb's translational structure the rectangular-bin choice used
in the main text is essentially the unique unitary window.

The most general linear decomposition of $\hat{a}[\omega]$ into a
countable family of baseband operators $\hat{a}_n[\Omega]$ is
\begin{align}
    \hat{a}_n[\Omega] &= \int_0^\infty \frac{d\omega}{2\pi}\,
        f_n[\omega,\Omega]\,\hat{a}[\omega],
        \label{eq:SI_fwd}\\
    \hat{a}[\omega]  &= \sum_n \int_{-\infty}^{\infty} \frac{d\Omega}{2\pi}\,
        g_n[\omega,\Omega]\,\hat{a}_n[\Omega],
        \label{eq:SI_inv}
\end{align}
where $\{f_n\}$ and $\{g_n\}$ are independent complex-valued kernel
families. For $\hat{a}[\omega]$ and $\{\hat{a}_n[\Omega]\}$ to describe
the same physical field, the map \cref{eq:SI_fwd} must be a unitary
change of basis. This requires two conditions on the kernel pair
$(f_n,g_n)$: surjectivity, that the inverse map \cref{eq:SI_inv} fully
reconstructs $\hat{a}[\omega]$; and isometry, that the forward map
\cref{eq:SI_fwd} preserves inner products.

Substituting \cref{eq:SI_fwd} into \cref{eq:SI_inv} and demanding
recovery of $\hat{a}[\omega]$ for arbitrary input gives the surjectivity
condition
\begin{equation}
    \sum_n \int_{-\infty}^{\infty} \frac{d\Omega}{2\pi}\,
    g_n[\omega,\Omega]\, f_n[\omega',\Omega]
    = 2\pi\,\delta[\omega-\omega'].
    \label{eq:SI_surj}
\end{equation}
Isometry is the orthonormality of the kernel family
$\{f_n[\cdot,\Omega]\}$ in the $\omega$ variable,
\begin{equation}
    \int_0^\infty \frac{d\omega}{2\pi}\,
    f_n[\omega,\Omega]^*\, f_m[\omega,\Omega']
    = 2\pi\,\delta_{nm}\,\delta[\Omega-\Omega'].
    \label{eq:SI_iso}
\end{equation}
Substituting \cref{eq:SI_fwd} into the commutator and using
\cref{eq:comm_delta},
\begin{align}
    [\hat{a}_n[\Omega],\,\hat{a}_m[\Omega']^\dagger]
    &= \int_0^\infty\!\frac{d\omega}{2\pi}\int_0^\infty\!\frac{d\omega'}{2\pi}\,
        f_n[\omega,\Omega]\, f_m[\omega',\Omega']^*\,
        [\hat{a}[\omega],\hat{a}[\omega']^\dagger] \notag\\
    &= \int_0^\infty \frac{d\omega}{2\pi}\,
        f_n[\omega,\Omega]\, f_m[\omega,\Omega']^*,
\end{align}
so isometry \cref{eq:SI_iso} is equivalent to the canonical bosonic
commutation relation
\begin{equation}
    [\hat{a}_n[\Omega],\,\hat{a}_m[\Omega']^\dagger]
    = 2\pi\,\delta_{nm}\,\delta[\Omega-\Omega'].
    \label{eq:SI_comm_target}
\end{equation}
Physically, isometry is the statement that the $\hat{a}_n[\Omega]$ are
mutually independent bosonic modes labelled by $(n,\Omega)$.

\Cref{eq:SI_surj,eq:SI_iso} admit many dual pairs $(f_n,g_n)$. The
simplest and most physically transparent choice is the self-dual one,
\begin{equation}
    g_n[\omega,\Omega] = f_n[\omega,\Omega]^*,
    \label{eq:SI_selfdual}
\end{equation}
for which the inverse map \cref{eq:SI_inv} is the adjoint of the forward
map \cref{eq:SI_fwd}. We adopt the self-dual form from here onwards.
Under \cref{eq:SI_selfdual}, surjectivity \cref{eq:SI_surj} becomes
\begin{equation}
    \sum_n \int_{-\infty}^{\infty} \frac{d\Omega}{2\pi}\,
    f_n[\omega,\Omega]^*\, f_n[\omega',\Omega]
    = 2\pi\,\delta[\omega-\omega'],
    \label{eq:SI_complete}
\end{equation}
which together with isometry \cref{eq:SI_iso} expresses unitarity of the
change of basis purely in terms of the single kernel family
$\{f_n[\omega,\Omega]\}$.

The comb's translational structure in frequency suggests restricting
attention to kernels that are covariant under shifts by the repetition
rate. A physically motivated ansatz is the windowed form
\begin{equation}
    f_n[\omega,\Omega]
    = 2\pi\,\delta[\omega - \omega_n - \Omega]\,f[\Omega],
    \qquad \omega_n = \omega_0 + n\Omega_r,
    \label{eq:SI_windowed}
\end{equation}
which reduces \cref{eq:SI_fwd} to
\begin{equation}
    \hat{a}_n[\Omega] = f[\Omega]\,\hat{a}[\omega_n+\Omega].
    \label{eq:SI_an_window}
\end{equation}
The translational ansatz reduces the entire kernel family to a single
window function $f[\Omega]$, and the two unitarity conditions become
explicit conditions on $f$.

Substituting \cref{eq:SI_windowed} into surjectivity \cref{eq:SI_complete}
gives the partition of unity
\begin{equation}
    \sum_n |f[\Omega - n\Omega_r]|^2 = 1,
    \label{eq:SI_POU}
\end{equation}
and the reconstruction formula takes the explicit form
\begin{equation}
    \hat{a}[\omega]
    = \sum_n f[\omega-\omega_n]^*\,\hat{a}_n[\omega-\omega_n].
    \label{eq:SI_recon}
\end{equation}
For the isometry side, the derived commutator of $\hat{a}_n[\Omega]$
follows from \cref{eq:SI_an_window,eq:comm_delta} as
\begin{equation}
    [\hat{a}_n[\Omega],\,\hat{a}_m[\Omega']^\dagger]
    = 2\pi\, f[\Omega]\, f[\Omega']^*\,
       \delta\!\bigl[(n-m)\Omega_r + \Omega - \Omega'\bigr],
    \label{eq:SI_comm_general}
\end{equation}
which matches the canonical form \cref{eq:SI_comm_target} exactly when
$|f[\Omega]| = 1$ on the support of $f$ and the support of $f$ and its
shift by $(m-n)\Omega_r$ are disjoint for all $n \neq m$. Together with
the partition of unity \cref{eq:SI_POU}, these two conditions force
$|f[\Omega]| = 1$ on an interval of width $\Omega_r$ and zero elsewhere.

The canonical choice is the rectangular bin
\begin{equation}
    f[\Omega] =
    \begin{cases}
        1 & -\Omega_r/2 \le \Omega < \Omega_r/2, \\
        0 & \text{otherwise},
    \end{cases}
    \label{eq:SI_rect}
\end{equation}
which we adopt from here onwards. With \cref{eq:SI_rect}, the partition
of unity is realised without tapering and without oversampling, so the
time--frequency lattice density sits at the critical Nyquist--Gabor value
$T_r\Omega_r = 2\pi$, saturating the Fourier-uncertainty bound on
phase-space sampling. The rectangular bin is moreover optimal in the
sense of Gabor localisation: by the Balian--Low theorem~\cite{Groch01},
at this critical density no unitary decomposition can be built from a
window that is simultaneously smooth in time and compactly supported in
frequency, so no alternative window yields a strictly better-localised
unitary decomposition. Under \cref{eq:SI_rect} the baseband operator
reduces to
\begin{equation}
\hat{a}_n[\Omega] = \hat{a}[\omega_n+\Omega],
\qquad -\Omega_r/2 \le \Omega < \Omega_r/2,
\label{eq:SI_an_rect}
\end{equation}
so that $\hat{a}_n[\Omega]$ is literally the comb-line-resolved field operator at offset $\Omega$ from the $n$-th comb line $\omega_n$, as used throughout the main
text. Specialising the derived commutator \cref{eq:SI_comm_general} to
\cref{eq:SI_rect} confirms the canonical form
\begin{equation}
    [\dop{a}_n[\Omega],\,\dop{a}_m[\Omega']^\dagger]
    = 2\pi\,\delta_{nm}\,\delta[\Omega-\Omega'],
    \label{eq:SI_comm_per_bin}
\end{equation}
for $\Omega,\Omega' \in [-\Omega_r/2,\Omega_r/2]$. This independence is what allows
us to assign quantum resources (coherent amplitudes, squeezing, EPR
entanglement) comb-line by comb-line in \cref{sec:SQL_OFD,sec:dualcomb}.

The time-domain baseband operator is recovered by inverse Fourier
transform,
\begin{equation}
    \hat{a}_n(t)
    = \int_{-\Omega_r/2}^{\Omega_r/2} \frac{d\Omega}{2\pi}\,
    \hat{a}_n[\Omega]\,e^{-i\Omega t},
\end{equation}
so that
\begin{equation}
    \hat{a}(t) = \sum_n \hat{a}_n(t)\,e^{-i\omega_n t},
    \qquad
    \hat{a}(t) = \sum_n\bigl[\alpha_n + \dop{a}_n(t)\bigr]\,e^{-i\omega_n t},
\end{equation}
which is the decomposition used throughout the main text.

\section{Optical frequency division}\label{sec:OFD_SI}

We derive the quantum noise floor of OFD through direct photodetection of the pulse train. We express the optical field operator as $\hat{a}(t) = \sum_n \left( \alpha_n + \delta\hat{a}_n(t) \right) e^{-in\Omega_r t}$, where $\alpha_n = |\alpha_n|e^{i\theta_n}$ is the coherent amplitude of the $n$-th comb line and $\delta\hat{a}_n(t)$ represents its quantum fluctuations. Assuming an ideal photodetector ($\eta=1$), the full photocurrent operator $\hat{I}(t)=q_e \hat{a}^\dagger(t)\hat{a}(t)$ expands to:
\begin{align}
\hat{I}(t)
&= q_e \sum_{n,n'} \alpha_{n'}^* \alpha_n\, e^{-i(n - n')\Omega_r t} \notag \\
&\quad + q_e \sum_{n,n'} \left[ \alpha_{n'}^*\, \delta\hat{a}_n(t) + \alpha_n \,\delta\hat{a}_{n'}^\dagger(t) \right] e^{-i(n - n')\Omega_r t} \notag \\
&\quad + q_e \sum_{n,n'} \delta\hat{a}_{n'}^\dagger(t)\, \delta\hat{a}_n(t)\, e^{-i(n - n')\Omega_r t}.
\label{eq:nt}
\end{align}
The first term represents the classical coherent beatnotes at harmonics of $\Omega_r$. The second term describes the quantum vacuum fluctuations amplified by the local oscillator power of the comb lines. The third term, purely vacuum noise, is vanishingly small and thus discarded.

We isolate the fundamental microwave beatnote at $\Omega_r$ by projecting $\hat{I}(t)$ onto the $e^{\mp i\Omega_r t}$ harmonics. Defining the positive-frequency component of the photocurrent at $\Omega_r$, we obtain:
\begin{equation}
\hat{I}_{\Omega_r}^{(+)}(t) = q_e \sum_n \left[
\alpha_{n-1}^* \alpha_n + \alpha_{n-1}^*\, \delta\hat{a}_n(t) + \alpha_{n+1}\, \delta\hat{a}_{n}^\dagger(t)
\right] e^{-i\Omega_r t},
\end{equation}
such that the physical harmonic operator is $\hat{I}_{\Omega_r}(t) = \hat{I}_{\Omega_r}^{(+)}(t) + \text{H.c.}$. 
Linearizing this operator around its coherent mean $\langle \hat{I}_{\Omega_r}^{(+)}(t)\rangle = q_e \sum_n \alpha_{n-1}^* \alpha_n \, e^{-i\Omega_r t} \equiv |S|e^{i\phi_0}e^{-i\Omega_r t}$, the fractional amplitude and phase noise estimators of the microwave carrier are defined via
\begin{equation}
    \hat{I}_{\Omega_r}^{(+)}(t) \approx |S|e^{i\phi_0}e^{-i\Omega_r t}\bigl(1 + \delta\tilde{A} + i\,\delta\tilde{\phi}\bigr) + O(\delta\hat{a}^2).
    \label{eq:linearization}
\end{equation}
The tildes in $\delta\tilde{A}$ and $\delta\tilde{\phi}$ signal that these are estimators defined through the first-order expansion of $\hat{I}_{\Omega_r}^{(+)}$, not intrinsic optical-field operators: the exact microwave phase would require $\arg[\hat{I}_{\Omega_r}^{(+)}]$, a nonlinear functional with no closed-form linear representation.
Both estimators are nonetheless well-defined Hermitian operators---linear combinations of the optical quadratures $\{\delta\hat{q}_n,\delta\hat{p}_n\}$ with real coefficients---and coincide with what a homodyne receiver phase-locked to the microwave beatnote actually measures.
Writing the quantum fluctuations in the quadrature basis $\delta \hat{q}_n \coloneqq (\delta \hat{a}_n + \delta \hat{a}_n^\dagger)/\sqrt{2}$ and $\delta \hat{p}_n \coloneqq (\delta \hat{a}_n - \delta \hat{a}_n^\dagger)/i\sqrt{2}$, we obtain:
\begin{align}
\delta\tilde{A}(t) &= \frac{1}{\sqrt{2}|S|} \sum_n \left[ \left( |\alpha_{n-1}| \cos\Delta\theta_n^{(-)} + |\alpha_{n+1}| \cos\Delta\theta_n^{(+)} \right) \delta\hat{q}_n(t) - \left( |\alpha_{n-1}| \sin\Delta\theta_n^{(-)} - |\alpha_{n+1}| \sin\Delta\theta_n^{(+)} \right) \delta\hat{p}_n(t) \right], \\
\delta\tilde{\phi}(t) &= \frac{1}{\sqrt{2}|S|} \sum_n \left[ \left( |\alpha_{n-1}| \sin\Delta\theta_n^{(-)} + |\alpha_{n+1}| \sin\Delta\theta_n^{(+)} \right) \delta\hat{q}_n(t) + \left( |\alpha_{n-1}| \cos\Delta\theta_n^{(-)} - |\alpha_{n+1}| \cos\Delta\theta_n^{(+)} \right) \delta\hat{p}_n(t) \right],
\end{align}
where $\Delta\theta_n^{(-)} \coloneqq \theta_{n-1} + \phi_0$ and $\Delta\theta_n^{(+)} \coloneqq \theta_{n+1} - \phi_0$.

To elucidate the physical consequence of these fluctuations, we consider an in-phase mode-locked comb where all comb lines share a uniform phase ($\alpha_n \in \mathbb{R}$, $\theta_n = 0$). Here, the temporal pulse profile is entirely governed by the spectral envelope $\{\alpha_n\}$. The amplitude and phase quadratures of the fundamental microwave harmonic simplify to:
\begin{equation}\label{eq:OFD_phase1}
\begin{aligned}
    \delta\tilde{A}(t) &= \frac{1}{\sqrt{2}|S|} \sum_{n=-N}^{N} \left( \alpha_{n-1} + \alpha_{n+1} \right) \delta\hat{q}_n(t), \\
    \delta\tilde{\phi}(t) &= \frac{1}{\sqrt{2}|S|} \sum_{n=-N}^{N} \left( \alpha_{n-1} - \alpha_{n+1} \right) \delta\hat{p}_n(t).
\end{aligned}
\end{equation}
Equation (\ref{eq:OFD_phase1}) reveals that the phase fluctuation $\delta\tilde{\phi}(t)$ is transduced from the phase quadrature $\delta\hat{p}_n$ of each line, weighted by the discrete spectral derivative $\alpha_{n-1} - \alpha_{n+1}$. This local spectral slope determines the sensitivity of the microwave timing jitter to individual per-line fluctuations.

\subsection{Effect of spectral shape on the standard quantum limit of OFD}\label{sec:SQL_OFD}

To systematically evaluate how the spectral envelope $\{\alpha_n\}$
shapes the quantum-limited phase noise of OFD, we first establish a
natural benchmark and a universal figure of merit.
The standard quantum limit (SQL) of the simplest possible microwave
generation scheme is set by a two-mode heterodyne measurement between
two continuous-wave (CW) lasers sharing a total photon flux
$N_\mathrm{tot} = P/\hbar\omega_0$.
Substituting $\alpha_{-1} = \alpha_0 = \sqrt{N_\mathrm{tot}/2}$,
$\alpha_n = 0$ otherwise into Eq.~\eqref{eq:OFD_phase1} yields:
\begin{equation}
    \bar{S}_{\tilde{\phi}\tilde{\phi}}^\mathrm{cw}
    = \frac{1}{N_\mathrm{tot}}.
    \label{eq:SQL_CW}
\end{equation}
To enable a physically meaningful comparison across comb spectra of
different shapes at fixed total optical power, we define a universal
root-mean-square (RMS) modal bandwidth,
\begin{equation}
    M_\mathrm{rms} = \sqrt{\frac{\sum_n n^2\alpha_n^2}{\sum_n \alpha_n^2}},
    \label{eq:Mrms_SI}
\end{equation}
which weights each line by its squared distance from the carrier and
provides a single scalar measure of effective spectral extent regardless
of envelope shape.
The normalized suppression ratio relative to the CW benchmark is then:
\begin{equation}
    R \equiv
    \bar{S}_{\tilde{\phi}\tilde{\phi}} \,/\,
    \bar{S}_{\tilde{\phi}\tilde{\phi}}^\mathrm{cw}
    = N_\mathrm{tot}\,\bar{S}_{\tilde{\phi}\tilde{\phi}},
\end{equation}
such that $R = 1$ recovers the CW SQL and $R < 1$ indicates genuine
improvement over heterodyne detection.

Assuming comb-line-resolved vacuum fluctuations that are spectrally
uncorrelated, $\bar{S}_{p_n p_n}[\Omega] = 1/2$ for each line, and
substituting into Eq.~\eqref{eq:OFD_phase1} yields the phase noise PSD
of the microwave carrier:
\begin{equation}
    \bar{S}_{\tilde{\phi}\tilde{\phi}}[\Omega]
    = \frac{1}{4|S|^2}\sum_{n=-N}^{N}
    \left(\alpha_{n-1} - \alpha_{n+1}\right)^2,
    \qquad |\Omega| \leq \frac{\Omega_r}{2},
    \label{eq:OFD_PSD}
\end{equation}
and zero otherwise.
The white, flat spectrum reflects the spectrally uncorrelated nature of
the vacuum fluctuations across distinct comb lines, while the discrete
spectral derivative $(\alpha_{n-1} - \alpha_{n+1})$ encodes the
complete dependence of the quantum noise floor on the envelope shape.

We evaluate $R$ for three archetypal spectral envelopes that collectively
represent the most technologically relevant classes of frequency comb
sources.
The first is a \textit{Gaussian} envelope,
$\alpha_n = A\exp(-n^2/4\sigma^2)$, characteristic of passively
mode-locked solid-state and fiber lasers whose gain bandwidth and
intracavity dispersion conspire to produce a smooth, bell-shaped spectrum.
The second is a \textit{sech} envelope,
$\alpha_n = A\,\mathrm{sech}(n/\Delta)$, which is the hallmark spectral
profile of dissipative Kerr solitons generated in
microresonators~\cite{Herr14,Kippenberg18} and soliton fiber lasers~\cite{HasegawaTappert73,Mollenauer80}; its heavier
tails relative to the Gaussian reflect the balance between anomalous
dispersion and Kerr nonlinearity that sustains the soliton.
The third is a \textit{flat-top} envelope, $\alpha_n = \alpha$ for
$|n| \leq N$ and zero otherwise, which represents the idealized limit of
a spectrally uniform comb and serves as the analytically tractable
limiting case that isolates the role of spectral boundaries.
These three shapes are deliberately chosen to span the space from smooth
and rapidly decaying (Gaussian) to smooth and slowly decaying (sech) to
abruptly truncated (flat-top), probing how the discrete spectral
derivative in Eq.~\eqref{eq:OFD_PSD} responds to different envelope
morphologies.
For each envelope, we analytically invert the exact relations
$M_\mathrm{rms}(\sigma)$, $M_\mathrm{rms}(\Delta)$, and
$M_\mathrm{rms}(N)$ to obtain the shape parameter corresponding to each
target $M_\mathrm{rms}$, construct $\{\alpha_n\}$ accordingly, and
evaluate $R$ exactly from Eq.~\eqref{eq:OFD_PSD}.

The numerical evaluation of $R$ for these three envelopes (see Fig.~\ref{fig:OFD}) confirms quadratic suppression $R\propto M_\mathrm{rms}^{-2}$ for smooth envelopes (Gaussian, sech) and only linear suppression $R\propto M_\mathrm{rms}^{-1}$ for flat-top, as discussed in the main text.

\subsection{Quantum-enhanced optical frequency division}
The phase-noise analysis of the preceding section assumed that the
comb-line-resolved field $\hat{a}_n[\Omega]$ is in the vacuum state for
every $n$ and $\Omega$, establishing the SQL as the baseline.
We now evaluate two distinct continuous-variable quantum strategies that
exploit non-classical correlations in $\{\delta\hat{p}_n\}$ to suppress
the phase noise below this limit.
The starting point is the fundamental microwave phase fluctuation
operator derived in \cref{eq:OFD_phase1}:
\begin{equation}
    \delta\tilde{\phi}(t)
    = \frac{1}{\sqrt{2}\,|S|}
      \sum_{n=-\infty}^{\infty}
      \left(\alpha_{n-1} - \alpha_{n+1}\right)\delta\hat{p}_n(t),
    \label{eq:OFD_phase_fluct}
\end{equation}
where $|S| = \sum_n \alpha_{n-1}\alpha_n$ is the macroscopic beatnote
amplitude and $\delta\hat{p}_n$ is the phase quadrature fluctuation
of the $n$-th comb line.
Because $\delta\tilde{\phi}$ is a weighted linear combination of the
phase quadratures $\{\delta\hat{p}_n\}$ with spectral-derivative
weights $(\alpha_{n-1} - \alpha_{n+1})$, any quantum state that reduces
the variance of this particular linear combination will yield a
sub-SQL phase noise floor.

\subsubsection*{Independent intra-comb-line squeezing}

The most direct approach is to prepare the comb-line-resolved field around every comb line in a phase-quadrature squeezed state, replacing the vacuum noise
$\bar{S}_{p_n p_n} = 1/2$ with
\begin{equation}
    \bar{S}_{p_n p_n} = \frac{1}{2G_n}, \qquad G_n > 1,
\end{equation}
where $G_n$ is the power squeezing factor for the $n$-th comb line.
Since the modes are independently squeezed, the phase noise PSD
of Eq.~\eqref{eq:OFD_PSD} becomes:
\begin{equation}
    \bar{S}_{\tilde{\phi}\tilde{\phi}}^{\mathrm{sq}}[\Omega]
    = \frac{1}{4|S|^2}
      \sum_{n=-N}^{N}
      \frac{\left(\alpha_{n-1} - \alpha_{n+1}\right)^2}{G_n},
    \label{eq:OFD_PSD_sq}
\end{equation}
which reduces to Eq.~\eqref{eq:OFD_PSD} when $G_n = 1$ for all $n$.
The improvement over the vacuum-state comb with the same spectral
envelope is characterised by the ratio:
\begin{equation}
    \eta_{\mathrm{sq}}
    \equiv
    \frac{\bar{S}_{\tilde{\phi}\tilde{\phi}}^{\mathrm{sq}}}
         {\bar{S}_{\tilde{\phi}\tilde{\phi}}^{\rm vac}}
    = \frac{\displaystyle\sum_{n=-N}^{N}
            \dfrac{(\alpha_{n-1}-\alpha_{n+1})^2}{G_n}}
           {\displaystyle\sum_{n=-N}^{N}
            (\alpha_{n-1}-\alpha_{n+1})^2},
    \label{eq:eta_sq}
\end{equation}
such that $\eta_{\mathrm{sq}} = 1$ recovers the vacuum SQL of the comb
and $\eta_{\mathrm{sq}} < 1$ quantifies the genuine quantum enhancement
attributable to squeezing alone, independent of any classical
advantage conferred by spectral broadening.
The combined suppression below the CW SQL is then simply
$R_{\mathrm{sq}} = R \cdot \eta_{\mathrm{sq}}$, cleanly separating the
classical contribution $R$ from the quantum contribution
$\eta_{\mathrm{sq}}$ (non-classical state preparation).

Equation~\eqref{eq:eta_sq} is the weighted harmonic mean of $1/G_n$ with squared spectral-derivative weights, so modes at steeper envelope slopes benefit most from stronger squeezing. For uniform squeezing $G_n = G$, $\eta_\mathrm{sq} = 1/G$ and the combined suppression is $R_\mathrm{sq} = R/G$.

\subsubsection*{Cross-comb-line entanglement}

A more resource-efficient approach exploits quantum correlations
between pairs of comb lines rather than within each line
individually.
Consider a comb with a symmetric spectral envelope,
$\alpha_n = \alpha_{-n}$, which is satisfied by any envelope
centred on the carrier — including all three archetypes studied in
Sec.~\ref{sec:SQL_OFD}.
Under this symmetry the spectral-derivative weights satisfy
\begin{equation}
    \alpha_{n-1} - \alpha_{n+1} = -(\alpha_{-n-1} - \alpha_{-n+1}),
    \label{eq:antisymmetry}
\end{equation}
so that the weight of line $+n$ is equal and opposite to that of
line $-n$.
In particular, the contribution of the central carrier line $n = 0$
vanishes exactly, since $\alpha_{-1} - \alpha_1 = 0$ by symmetry.

Exploiting Eq.~\eqref{eq:antisymmetry}, the sum in
Eq.~\eqref{eq:OFD_phase_fluct} can be folded around the carrier.
Grouping the $+n$ and $-n$ contributions for $n \geq 1$:
\begin{align}
    \delta\tilde{\phi}(t)
    &= \frac{1}{\sqrt{2}\,|S|}
       \sum_{n=1}^{\infty}
       \left(\alpha_{n-1} - \alpha_{n+1}\right)
       \left[\delta\hat{p}_n(t) - \delta\hat{p}_{-n}(t)\right].
    \label{eq:OFD_folded}
\end{align}
This form reveals that the microwave phase fluctuation depends
exclusively on the antisymmetric phase-quadrature combinations
$\delta\hat{p}_n - \delta\hat{p}_{-n}$; the symmetric combinations
$\delta\hat{p}_n + \delta\hat{p}_{-n}$ are entirely decoupled from
the phase and contribute nothing to the phase noise.

It is natural to introduce the antisymmetric EPR phase quadrature
operator for the $n$-th comb-line pair:
\begin{equation}
    \hat{P}_n^-(t)
    \coloneqq \frac{\delta\hat{p}_n(t) - \delta\hat{p}_{-n}(t)}{\sqrt{2}},
    \label{eq:EPR_phase}
\end{equation}
normalised so that $\bar{S}_{P_n^- P_n^-} = 1/2$ when both modes are
in the vacuum.
Substituting Eq.~\eqref{eq:EPR_phase} into Eq.~\eqref{eq:OFD_folded},
the phase fluctuation collapses into a sum over EPR quadratures alone:
\begin{equation}
    \delta\tilde{\phi}(t)
    = \frac{1}{|S|}
      \sum_{n=1}^{\infty}
      \left(\alpha_{n-1} - \alpha_{n+1}\right)\hat{P}_n^-(t).
    \label{eq:OFD_EPR}
\end{equation}
The SQL is recovered when $\bar{S}_{P_n^- P_n^-} = 1/2$ for all $n$,
consistent with Eq.~\eqref{eq:OFD_PSD} after accounting for the
factor of two from folding.

Equation~\eqref{eq:OFD_EPR} is the central result of this section.
It shows that cross-comb-line entanglement---specifically, squeezing of
the antisymmetric EPR phase quadrature $\hat{P}_n^-$ for each
symmetrically placed pair $(+n, -n)$---constitutes a sufficient
quantum resource for sub-SQL OFD, complementary to the individual
comb-line squeezing considered previously.
The symmetric EPR phase quadrature $\hat{P}_n^+ \coloneqq
(\delta\hat{p}_n + \delta\hat{p}_{-n})/\sqrt{2}$ is completely
invisible to the OFD phase measurement and may therefore have
arbitrary quantum statistics.
Allowing a pair-dependent squeezing factor $G_n > 1$ such that
$\bar{S}_{P_n^- P_n^-} = 1/(2G_n)$, the phase noise PSD becomes:
\begin{equation}
    \bar{S}_{\tilde{\phi}\tilde{\phi}}^{\mathrm{ent}}[\Omega]
    = \frac{1}{|S|^2}
      \sum_{n=1}^{\infty}
      \frac{\left(\alpha_{n-1} - \alpha_{n+1}\right)^2}{2G_n},
\end{equation}
and the improvement over the vacuum-state comb is characterised
by the same weighted harmonic mean structure as intra-comb-line squeezing:
\begin{equation}
    \eta_{\mathrm{ent}}
    \equiv
    \frac{\bar{S}_{\tilde{\phi}\tilde{\phi}}^{\mathrm{ent}}}
         {\bar{S}_{\tilde{\phi}\tilde{\phi}}^{\rm vac}}
    = \frac{\displaystyle\sum_{n=1}^{\infty}
            \dfrac{(\alpha_{n-1}-\alpha_{n+1})^2}{G_n}}
           {\displaystyle\sum_{n=1}^{\infty}
            (\alpha_{n-1}-\alpha_{n+1})^2}.
    \label{eq:eta_EPR}
\end{equation}
Importantly, symmetric signal--idler sideband pairs are the natural quantum
degrees of freedom produced in parametric down-conversion and in vacuum-seeded four-wave mixing in Kerr microresonators~\cite{Chembo16,Roslund14}.
The resulting two-mode correlations suggest a possible route toward intracavity quantum enhancement without an external squeezing apparatus.

\subsection{Classical noise limit of optical frequency division}
\label{sec:OFD_classical}

We demonstrate that the comb-line-resolved operator formalism rigorously 
recovers the classical limit of optical frequency division. Consider a 
frequency comb whose central line is tightly phase-locked to a classical 
optical reference, such that its frequency fluctuation tracks the reference, 
$\delta\omega_0(t) = \delta\omega_{\mathrm{ref}}(t)$. The comb carrier 
frequency is parameterized as $\omega_0 = \Omega_c + N_0\Omega_r$, where
$N_0 \gg 1$ is the line index of the optical carrier and the carrier-envelope offset, $\Omega_c$, is rigidly stabilized via $f$-$2f$ interferometry.
The frequency fluctuations of the repetition rate are then strictly coupled to the reference via 
$\delta\Omega_r(t) = \delta\omega_{\mathrm{ref}}(t)/N_0$.

The frequency fluctuation of the $n$-th comb line, 
$\omega_n = \omega_0 + n\Omega_r$, is therefore:
\begin{equation}
    \delta\omega_n(t) = \delta\omega_0(t) + n\,\delta\Omega_r(t) 
    = \left(1 + \frac{n}{N_0}\right)\delta\omega_{\mathrm{ref}}(t).
\end{equation}
Since phase is the time integral of frequency, the classical phase 
fluctuation of the $n$-th comb line scales identically:
\begin{equation}
    \delta\phi_n(t) = \left(1 + \frac{n}{N_0}\right)\delta\phi_{\mathrm{ref}}(t).
\end{equation}
Within the linearized quadrature operator formalism, a pure classical 
phase fluctuation can be represented as phase quadrature fluctuations of the form $\delta\hat{p}_n(t) = \sqrt{2}\alpha_n\,\delta\phi_n(t) + \delta\hat{p}_n^0(t)$, where $\alpha_n$ is the coherent amplitude of the $n$-th comb line and $\delta\hat{p}_n^0(t)$ represents the vacuum contribution. 
Substituting into Eq.~\eqref{eq:OFD_phase} of the main text:
\begin{equation}
    \delta\tilde{\phi}(t) = \frac{1}{\sqrt{2}|S|} \sum_{n}
    \left(\alpha_{n-1} - \alpha_{n+1}\right)
    \left[\sqrt{2}\alpha_n\left(1 + \frac{n}{N_0}\right)
    \delta\phi_{\mathrm{ref}}(t)\right],
\end{equation}
where $|S| = \sum_n \alpha_{n-1}\alpha_n$.
Here we have neglected the vacuum contribution $\delta\hat{p}_n^0(t)$.
Separating the sum and shifting the summation index $n \to n-1$ in the
second term:
\begin{align}
    \delta\tilde{\phi}(t)
    &= \frac{\delta\phi_{\mathrm{ref}}(t)}{|S|}
    \Bigg[
    \sum_{n}\alpha_{n-1}\alpha_n\left(1+\frac{n}{N_0}\right)
    - \sum_{n}\alpha_{n+1}\alpha_n\left(1+\frac{n}{N_0}\right)
    \Bigg] \nonumber \\
    &= \frac{\delta\phi_{\mathrm{ref}}(t)}{|S|} \sum_{n}
    \alpha_{n-1}\alpha_n
    \left[\left(1+\frac{n}{N_0}\right)
    - \left(1+\frac{n-1}{N_0}\right)\right].
\end{align}
The bracketed term evaluates exactly to $1/N_0$, and the remaining sum 
$\sum_n\alpha_{n-1}\alpha_n = |S|$ cancels the prefactor, yielding:
\begin{equation}
    \delta\tilde{\phi}(t) = \frac{\delta\phi_{\mathrm{ref}}(t)}{N_0}
    = \frac{\Omega_r}{\omega_0}\,\delta\phi_{\mathrm{ref}}(t).
    \label{eq:OFD_classical_SI}
\end{equation}
This result carries two important physical implications. First, the phase 
noise PSD of the extracted microwave signal is suppressed by 
$(\Omega_r/\omega_0)^2$ relative to the optical reference, exactly reproducing
the standard classical OFD suppression factor from first principles within 
the quantum operator formalism. Second, this result is entirely independent 
of the spectral envelope $\{\alpha_n\}$: the discrete spectral derivative 
nullifies the envelope dependence in the classical regime. This establishes 
a sharp physical dichotomy --- while the classical noise floor is immune 
to the comb's spectral shape, the quantum noise floor is sensitive 
to it.

\section{Dual-Comb Spectroscopy}\label{sec:dualcomb}

We analyze DCS in the asymmetric balanced-detection configuration, in which the signal comb $\hat{a}_S(t)$ passes through the sample and the reference comb $\hat{a}_L(t)$ serves as a local oscillator.
The two fields are combined on a 50:50 beam splitter and the output ports are measured by a balanced photodetector:
\begin{equation}
\begin{pmatrix}
\hat{a}_+(t) \\ \hat{a}_-(t)
\end{pmatrix}
=
\frac{1}{\sqrt{2}}
\begin{pmatrix}
1 & 1 \\ -1 & 1
\end{pmatrix}
\begin{pmatrix}
\hat{a}_S(t) \\ \hat{a}_L(t)
\end{pmatrix}.
\end{equation}
The balanced photocurrent
$\hat{I}(t) = q_e[\hat{a}_+^\dagger(t)\hat{a}_+(t)
- \hat{a}_-^\dagger(t)\hat{a}_-(t)]$
rigorously rejects the direct self-beating of each comb, nulling
the macroscopic common-mode intensity noise and leaving only the
cross-correlation:
\begin{equation}
    \hat{I}(t)
    = q_e\bigl(
        \hat{a}_S^\dagger(t)\hat{a}_L(t)
        + \hat{a}_L^\dagger(t)\hat{a}_S(t)
      \bigr).
    \label{eq:balanced_photocurrent}
\end{equation}

\subsection{Sample response model}

The interaction of the signal comb with the sample is modeled as a
linear, frequency-dependent transformation of the field annihilation
operator.
To preserve the canonical commutation relations under macroscopic
attenuation, any intensity loss must be accompanied by a proportional
coupling to environmental vacuum modes.
The sample response map $\mathcal{L}$ acting on the continuous-mode
operator $\hat{a}[\omega]$ is:
\begin{equation}
    \mathcal{L}\bigl(\hat{a}[\omega]\bigr)
    = \sqrt{\kappa[\omega]}\,e^{i\theta[\omega]}\,\hat{a}[\omega]
    + \sqrt{1-\kappa[\omega]}\,\hat{v}[\omega],
    \label{eq:sample_response}
\end{equation}
where $\kappa[\omega] \in [0,1]$ is the intensity transmittance,
$\theta[\omega]$ is the phase delay, and $\hat{v}[\omega]$ is an
uncorrelated vacuum mode satisfying
$[\hat{v}[\omega],\hat{v}[\omega']^\dagger]
= 2\pi\delta[\omega-\omega']$.

The multi-heterodyne beat notes are confined to a tight RF bandwidth
$M\,\Delta\Omega_r$ that is vastly smaller than the optical comb-line
spacing ($M\,\Delta\Omega_r \ll \Omega_r/2$).
The measurement therefore probes the sample response in an
infinitesimally narrow window around each comb-line centre
$\omega_n$, justifying a piecewise-constant approximation:
\begin{equation}
    \kappa[\omega_n+\Omega] \simeq \kappa_n,
    \qquad
    \theta[\omega_n+\Omega]  \simeq \theta_n,
    \qquad
    |\Omega| \leq M\,\Delta\Omega_r.
\end{equation}
The comb-line-resolved sample map then takes the form:
\begin{equation}
    \mathcal{L}\bigl(\hat{a}_n(t)\bigr)
    = \sqrt{\kappa_n}\,e^{i\theta_n}\,\hat{a}_n(t)
    + \sqrt{1-\kappa_n}\,\hat{v}_n(t),
    \label{eq:sample_response_discrete}
\end{equation}
mapping the broadband optical susceptibility onto a discrete set of
independent scattering channels, each carrying its own vacuum noise
penalty from optical loss.

\subsection{Transmittance estimator and signal-to-noise ratio}
\label{sec:SNR_kappa}
To connect the photocurrent PSDs in DCS to the practical figure of merit---the precision with which the sample transmittance $\kappa_m$ can be extracted---we derive the power signal-to-noise ratio (SNR) for $\kappa_m$ in terms of $\bar{S}_{II}[\Omega]$ and the measurement time~$T$.

The transmittance $\kappa_m$ is encoded in the amplitude of the RF beat note at $\Omega_m = \Omega_0 + m\,\Delta\Omega_r$.
To isolate this tone, we demodulate the photocurrent at frequency $\Omega_m$ with the matched phase $\theta_m$ and average over the measurement window~$T$:
\begin{equation}
    \tilde{I}_m
    = \frac{2}{T}\int_0^T \hat{I}(t)\,
      \cos(\Omega_m t - \theta_m)\,dt.
    \label{eq:demod_estimator}
\end{equation}
Substituting the mean photocurrent, which follows from Eq.~\eqref{eq:balanced_photocurrent} with the sample response Eq.~\eqref{eq:sample_response_discrete},
\begin{equation}
    \bigl\langle \hat{I}(t) \bigr\rangle
    = 2q_e \sum_n \sqrt{\kappa_n}\,\alpha_{S,n}\alpha_{L,n}
      \cos(\Omega_n t - \theta_n),
    \label{eq:mean_photocurrent_sample}
\end{equation}
the orthogonality of the RF tones over $T \gg 2\pi/\Delta\Omega_r$ selects a single beat note:
\begin{equation}
    \bigl\langle \tilde{I}_m \bigr\rangle
    = 2q_e\sqrt{\kappa_m}\,\alpha_{S,m}\alpha_{L,m}.
    \label{eq:demod_mean}
\end{equation}
Because this amplitude scales as $\sqrt{\kappa_m}$, a natural estimator for the transmittance is
\begin{equation}
    \hat{\kappa}_m
    = \frac{\tilde{I}_m^{\,2}}
           {4q_e^2\,\alpha_{S,m}^2\alpha_{L,m}^2},
    \label{eq:kappa_estimator}
\end{equation}
which requires independent knowledge of the comb amplitudes $\alpha_{S,m}$ and $\alpha_{L,m}$, obtainable from a reference measurement without the sample.

The variance of the demodulated estimator~\eqref{eq:demod_estimator} is
\begin{equation}
\begin{split}
    \mathrm{Var}(\tilde{I}_m)
    &= \frac{4}{T^2}
      \int_0^T\!\!\int_0^T
        \bar{C}_{II}(t-t')\\
    &\quad\times
        \cos(\Omega_m t - \theta_m)\,
        \cos(\Omega_m t' - \theta_m)\,
      dt\,dt',
\end{split}
    \label{eq:var_double_integral}
\end{equation}
where $\bar{C}_{II}(\tau) \equiv \tfrac{1}{2}\langle\{\delta\hat{I}(t+\tau),\,\delta\hat{I}(t)\}\rangle$ is the symmetrized autocorrelation, which depends only on the lag $\tau = t - t'$ by the (wide-sense) stationarity of the photocurrent noise.
This can be decomposed into a slowly varying term and a rapidly oscillating term at twice the carrier:
\begin{equation}
\begin{split}
    \mathrm{Var}(\tilde{I}_m)
    &= \frac{2}{T^2}
      \int_0^T\!\!\int_0^T
        \bar{C}_{II}(\tau)\\
    &\quad\times\bigl[
          \cos\Omega_m\tau
          + \cos\bigl(\Omega_m(t+t') - 2\theta_m\bigr)
        \bigr]\,dt\,dt',
\end{split}
    \label{eq:var_product_to_sum}
\end{equation}
for $\Omega_m T \gg 1$, the fast term averages to zero.
Changing variables to $\tau = t - t'$ and extending the integration limits (valid when $T$ greatly exceeds the correlation time of $\bar{C}_{II}$), the variance reduces to
\begin{equation}
\begin{split}
    \mathrm{Var}(\tilde{I}_m)
    &= \frac{2}{T}
      \int_{-\infty}^{\infty}
        \bar{C}_{II}(\tau)\,\cos\Omega_m\tau\,d\tau\\
    &= \frac{\bar{S}_{II}[\Omega_m] + \bar{S}_{II}[-\Omega_m]}{T}
     = \frac{2\bar{S}_{II}[\Omega_m]}{T},
\end{split}
    \label{eq:var_from_PSD}
\end{equation}
where we identified the integral as the (two-sided) symmetrized spectral density via the Wiener--Khinchin relation (Sec.~\ref{sec:spectral_densities}), and used $\bar{S}_{II}[\Omega] = \bar{S}_{II}[-\Omega]$.

Linearising the estimator~\eqref{eq:kappa_estimator} around its expectation value and propagating the variance gives
\begin{equation}
    \mathrm{Var}(\hat{\kappa}_m)
    = \left(\frac{\partial\hat{\kappa}_m}{\partial\tilde{I}_m}
      \right)^{\!2}
      \mathrm{Var}(\tilde{I}_m)
    = \frac{\kappa_m}
           {q_e^2\,\alpha_{S,m}^2\alpha_{L,m}^2}
      \cdot \frac{2\bar{S}_{II}[\Omega_m]}{T}.
    \label{eq:var_kappa}
\end{equation}
Defining the power signal-to-noise ratio as $\mathrm{SNR}_{\kappa_m} \equiv \kappa_m^2/\mathrm{Var}(\hat{\kappa}_m)$:
\begin{equation}
    \mathrm{SNR}_{\kappa_m}
    = \frac{\kappa_m\,q_e^2\,
            \alpha_{S,m}^2\,\alpha_{L,m}^2\,T}
           {2\,\bar{S}_{II}[\Omega_m]}.
    \label{eq:SNR_kappa}
\end{equation}
For fixed comb amplitudes and sample transmittance, the SNR is determined entirely by the ratio $T/\bar{S}_{II}$: any quantum protocol that reduces the photocurrent PSD yields a proportional improvement in the absorption measurement precision.
By inserting the SQL photocurrent PSD from Eq.~\eqref{eq:DCS_SQL} into Eq.~\eqref{eq:SNR_kappa}, we recover the SQL SNR in Eq.~\eqref{eq:DCS_SNR_SQL}.

\subsection{Self-referred intra-comb-line squeezing}
We begin with the self-referred case, which serves primarily
as a cautionary foil.
In the self-referred basis, quantum fluctuations are defined
relative to each comb's own native carrier frequencies.
The signal and reference combs are written as:
\begin{equation}
\begin{aligned}
\hat{a}_{S}(t) &= \sum_{n}
    \bigl[\alpha_{S,n} + \delta\hat{a}_{S,n}(t)\bigr]
    e^{-i(\omega_0 + n\Omega_r)t},\\
\hat{a}_{L}(t) &= \sum_{n}
    \bigl[\alpha_{L,n} + \delta\hat{a}_{L,n}(t)\bigr]
    e^{-i(\omega_0 + \Omega_0 + n(\Omega_r+\Delta\Omega_r))t}.
\end{aligned}
\label{eq:comb_self}
\end{equation}
The balanced photocurrent fluctuation from
Eq.~\eqref{eq:balanced_photocurrent}, retaining only terms within
the detection bandwidth, is:
\begin{equation}
\begin{aligned}
\delta\hat{I}(t)
= \sqrt{2}\,q_e \sum_n \Big[
&\bigl(\alpha_{S,n}\,\delta\hat{q}_{L,n}(t)
+ \alpha_{L,n}\,\delta\hat{q}_{S,n}(t)\bigr)\cos\Omega_n t \\
+\,&\bigl(\alpha_{S,n}\,\delta\hat{p}_{L,n}(t)
- \alpha_{L,n}\,\delta\hat{p}_{S,n}(t)\bigr)\sin\Omega_n t
\Big],
\end{aligned}
\label{eq:photocurrent_self}
\end{equation}
where $\delta\hat{q}$ and $\delta\hat{p}$ are the amplitude and
phase quadratures referenced to each comb's own frequency grid.

If both combs are prepared in independent squeezed states with
per-line squeezing gains $G_{S,n}$ and $G_{L,n}$ ($G = 1$ is
the vacuum limit), the quadrature PSDs of the $n$-th comb line are
$\bar{S}_{qq}^{S/L,n} = 1/(2G_{S/L,n})$ and
$\bar{S}_{pp}^{S/L,n} = G_{S/L,n}/2$.
Including the sample response, the photocurrent PSD becomes:
\begin{equation}
\bar{S}_{II}[\Omega]
= \frac{q_e^2}{2}\sum_n \Bigg[
\kappa_n\alpha_{S,n}^2\Bigl(G_{L,n} + \frac{1}{G_{L,n}}\Bigr)
+ \alpha_{L,n}^2\Bigl(
    \kappa_n\Bigl(G_{S,n} + \frac{1}{G_{S,n}}\Bigr)
    + 2(1-\kappa_n)
  \Bigr)
\Bigg].
\label{eq:PSD_self}
\end{equation}
The origin of the $G_{S/L,n} + 1/G_{S/L,n}$ factor is the cyclostationary nature
of the measurement: the native squeezed quadrature rotates
continuously relative to the multi-heterodyne beat notes at
frequency $\Omega_n$, so the detector time-averages both the
squeezed ($1/G_{S/L,n}$) and anti-squeezed ($G_{S/L,n}$) variances
with equal weight.
Since $G_n + 1/G_n \geq 2$ for any $G_n \geq 1$,
self-referred squeezing paradoxically \textit{increases} the noise
above the vacuum SQL, making conventional self-squeezed combs
fundamentally incompatible with heterodyne dual-comb spectroscopy
(see Fig.~\ref{fig:SI_cyclo} for a numerical illustration).

\subsection{Numerical simulation of cyclostationary noise}

\begin{figure}[t!]
    \centering
    \includegraphics[width=0.5\textwidth]{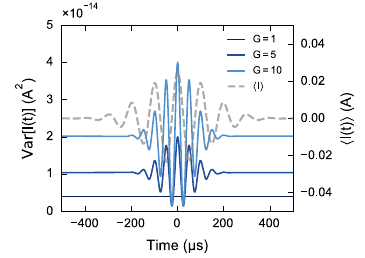}
    \caption{\textbf{Cyclostationary photocurrent variance for self-referred squeezing.}
    Time-domain photocurrent variance $\mathrm{Var}[I(t)]$ (left axis, solid lines) oscillates between the squeezed level $1/G$ and the anti-squeezed level $G$ for gains $G=1$ (dark blue), $G=5$ (medium blue), and $G=10$ (light blue), shown together with the mean photocurrent $\langle I(t)\rangle$ (right axis, gray dashed). The variance is up- and down-modulated at twice the beat frequencies $\Omega_n$, time-averaging to a noise floor that is increased by $(G+1/G)/2 \geq 1$ above the vacuum SQL.}
    \label{fig:SI_cyclo}
\end{figure}

The time-domain photocurrent traces in Fig.~\ref{fig:SI_cyclo} are generated by a direct numerical simulation of the self-referred squeezing scenario.
The signal comb has a center wavelength $\lambda_0 = 1550\,\mathrm{nm}$ and total power $P = 10\,\mathrm{mW}$, with $M = 101$ comb lines indexed $n = -50, \ldots, +50$.
The amplitude envelope follows a Gaussian profile $|\alpha_n| \propto \exp(-n^2 / 2\sigma^2)$ with $\sigma = N/3 \approx 16.67$ ($N = 50$), normalized to the total power.
Per-line phase-quadrature squeezing in the self-referred frame is applied independently to each line with gain $G \in \{1, 5, 10\}$.
The photocurrent operator $\hat{I}(t) = q_e \hat{a}^\dagger(t)\hat{a}(t)$ is evaluated in the strong-LO approximation [Eq.~\eqref{eq:DCS_photocurrent_LO}], and the instantaneous variance $\mathrm{Var}[I(t)]$ is computed analytically from the quadrature-operator expressions at each time step.
The simulation uses a sampling rate of $f_s = 1\,\mathrm{MHz}$ (time step $dt = 1\,\mu\mathrm{s}$) and a resolution bandwidth of $\Delta\Omega / 2\pi = 100\,\mathrm{Hz}$.

\subsection{Cross-referred intra-comb-line squeezing}
\label{sec:cross_referred}

The cyclostationary averaging is eliminated by redefining the
quantum fluctuation operators in the rotating frame of the
multi-heterodyne beat notes — that is, by referencing the
fluctuations of each comb to the frequency grid of the other.
Under the substitution
$\delta\hat{a}_{X,n}(t)
\to \delta\hat{a}_{X,n}(t)\,e^{i(\Omega_0 + n\Delta\Omega_r)t}$,
the comb fields become:
\begin{equation}
\begin{aligned}
\hat{a}_{S}(t)
&= \sum_n \bigl[
    \alpha_{S,n}\,e^{-i(\omega_0+n\Omega_r)t}
    + \delta\hat{a}_{S,n}(t)\,
      e^{-i(\omega_0+\Omega_0+n(\Omega_r+\Delta\Omega_r))t}
  \bigr],\\
\hat{a}_{L}(t)
&= \sum_n \bigl[
    \alpha_{L,n}\,e^{-i(\omega_0+\Omega_0+n(\Omega_r+\Delta\Omega_r))t}
    + \delta\hat{a}_{L,n}(t)\,e^{-i(\omega_0+n\Omega_r)t}
  \bigr].
\end{aligned}
\label{eq:comb_cross}
\end{equation}
In this cross-referred basis the beat notes perfectly demodulate the
quantum fluctuations, and the balanced photocurrent fluctuation
reduces to a manifestly non-cyclostationary form:
\begin{equation}
    \delta\hat{I}(t)
    = \sqrt{2}\,q_e \sum_n
      \bigl[
        \alpha_{S,n}\,\delta\hat{q}_{L,n}(t)
        + \alpha_{L,n}\,\delta\hat{q}_{S,n}(t)
      \bigr],
    \label{eq:photocurrent_cross}
\end{equation}
where only the amplitude quadratures $\delta\hat{q}$ contribute —
the phase quadratures $\delta\hat{p}$ are completely decoupled from
the photocurrent in this basis.
Including the sample response, the photocurrent PSD is:
\begin{equation}
\bar{S}_{II}^{\mathrm{sq}}[\Omega]
= q_e^2 \sum_n \Bigg[
    \frac{\kappa_n}{G_{L,n}}\,\alpha_{S,n}^2
    + \alpha_{L,n}^2\Bigl\{
        \kappa_n\Bigl(\frac{\cos^2\theta_n}{G_{S,n}}
                    + G_{S,n}\sin^2\theta_n\Bigr)
        + (1-\kappa_n)
      \Bigr\}
\Bigg].
\label{eq:PSD_cross}
\end{equation}
For a transparent, dispersion-free sample ($\kappa_n = 1$,
$\theta_n = 0$), the noise reduces to
$q_e^2\sum_n(\alpha_{S,n}^2/G_{L,n} + \alpha_{L,n}^2/G_{S,n})$,
genuinely surpassing the SQL.
However, Eq.~\eqref{eq:PSD_cross} reveals a fundamental
vulnerability: any non-zero sample phase delay $\theta_n$ rotates
the measurement quadrature, mixing the anti-squeezed variance
$G_{S,n}\sin^2\theta_n$ into the photocurrent.
Cross-referred squeezing therefore provides genuine quantum
enhancement only when the sample's phase dispersion is negligible
or actively pre-compensated.

\subsection{Cross-comb-line entanglement}

This approach exploits quantum correlations distributed
across symmetric pairs of comb lines $(+n, -n)$ within each comb
via a two-mode squeezing (TMS) interaction.
The natural language for this is the EPR quadrature basis.
For each pair index $n \geq 1$, we define the symmetric and
antisymmetric EPR quadratures:
\begin{equation}
    \hat{Q}_{J,n}^{\pm}(t)
    = \frac{\delta\hat{q}_{J,+n}(t) \pm \delta\hat{q}_{J,-n}(t)}{\sqrt{2}},
    \qquad
    \hat{P}_{J,n}^{\pm}(t)
    = \frac{\delta\hat{p}_{J,+n}(t) \pm \delta\hat{p}_{J,-n}(t)}{\sqrt{2}},
    \label{eq:EPR_def}
\end{equation}
for comb $J \in \{S, L\}$.
If the $(+n,-n)$ pair of comb $J$ is prepared in a two-mode
squeezed state with gain $G_{J,n} \geq 1$, the PSDs of the
EPR quadratures satisfy:
\begin{equation}
    \bar{S}^J_{Q_n^+ Q_n^+}
    = \bar{S}^J_{P_n^- P_n^-}
    = \frac{1}{2G_{J,n}},
    \qquad
    \bar{S}^J_{Q_n^- Q_n^-}
    = \bar{S}^J_{P_n^+ P_n^+}
    = \frac{G_{J,n}}{2},
    \label{eq:EPR_PSD}
\end{equation}
with vanishing cross-correlations between different EPR quadratures.

Under the symmetric-comb assumption
($\alpha_{S,n} = \alpha_{S,-n}$, $\alpha_{L,n} = \alpha_{L,-n}$),
regrouping the $\pm n$ pairs in Eq.~\eqref{eq:photocurrent_cross}
shows that the cross-referred photocurrent couples exclusively to
the symmetric EPR amplitude quadrature $\hat{Q}^+_{J,n}$:
\begin{equation}
    \delta\hat{I}(t)
    = \sqrt{2}\,q_e\,\bigl[
        \alpha_{S,0}\,\delta\hat{q}_{L,0}(t)
        + \alpha_{L,0}\,\delta\hat{q}_{S,0}(t)
      \bigr]
    + 2q_e \sum_{n=1}^{N}
      \bigl[
        \alpha_{S,n}\,\hat{Q}^+_{L,n}(t)
        + \alpha_{L,n}\,\hat{Q}^+_{S,n}(t)
      \bigr].
    \label{eq:photocurrent_EPR_ideal}
\end{equation}
Two-mode squeezing directly suppresses $\bar{S}^J_{Q_n^+ Q_n^+}$,
and the measurement is entirely immunised against the anti-squeezed
quadrature $\hat{Q}^-_{J,n}$ in this idealised, sample-free case.

In practice, a general sample breaks the symmetry between the $+n$
and $-n$ lines through unequal absorption ($\kappa_{+n} \neq \kappa_{-n}$)
or non-odd phase dispersion ($\theta_{+n} \neq -\theta_{-n}$).
Including the sample response on the signal comb and projecting onto
the EPR basis, the photocurrent contribution from the $n$-th pair
becomes:
\begin{equation}
\begin{aligned}
    \delta\hat{I}_n(t)
    = q_e \Big\{
    &\alpha_{S,n}\Big[
        \bigl(\sqrt{\kappa_{+n}}+\sqrt{\kappa_{-n}}\bigr)\hat{Q}^+_{L,n}
        + \bigl(\sqrt{\kappa_{+n}}-\sqrt{\kappa_{-n}}\bigr)\hat{Q}^-_{L,n}
      \Big]\\
    +\, &\alpha_{L,n}\Big[
        \bigl(\sqrt{\kappa_{+n}}\cos\theta_{+n}+\sqrt{\kappa_{-n}}\cos\theta_{-n}\bigr)\hat{Q}^+_{S,n}
        + \bigl(\sqrt{\kappa_{+n}}\cos\theta_{+n}-\sqrt{\kappa_{-n}}\cos\theta_{-n}\bigr)\hat{Q}^-_{S,n}\\
        &\quad+\bigl(\sqrt{\kappa_{+n}}\sin\theta_{+n}+\sqrt{\kappa_{-n}}\sin\theta_{-n}\bigr)\hat{P}^+_{S,n}
        + \bigl(\sqrt{\kappa_{+n}}\sin\theta_{+n}-\sqrt{\kappa_{-n}}\sin\theta_{-n}\bigr)\hat{P}^-_{S,n}\\
        &\quad+\bigl(\sqrt{1-\kappa_{+n}}+\sqrt{1-\kappa_{-n}}\bigr)\hat{Q}^+_{v,n}
        + \bigl(\sqrt{1-\kappa_{+n}}-\sqrt{1-\kappa_{-n}}\bigr)\hat{Q}^-_{v,n}
      \Big]
    \Big\}.
\end{aligned}
\label{eq:photocurrent_EPR_sample}
\end{equation}
The physical penalty is explicit: any sample-induced asymmetry in
either absorption or dispersion couples the photocurrent to the
anti-squeezed quadratures $\hat{Q}^-_{S,n}$, $\hat{P}^+_{S,n}$,
and $\hat{Q}^-_{L,n}$, whose variances grow in proportion to the
squeezing gain $G_{J,n}$.
The resulting photocurrent PSD for the $n$-th pair is:
\begin{equation}
\begin{aligned}
\bar{S}_{II,n}^{\mathrm{ent}}[\Omega]
= \frac{q_e^2}{2}
\Bigg\{
&\alpha_{S,n}^2\Bigg[
    (\kappa_{+n}+\kappa_{-n})\Bigl(G_{L,n}+\frac{1}{G_{L,n}}\Bigr)
    -2\sqrt{\kappa_{+n}\kappa_{-n}}\Bigl(G_{L,n}-\frac{1}{G_{L,n}}\Bigr)
  \Bigg]\\
+\,&\alpha_{L,n}^2\Bigg[
    (\kappa_{+n}+\kappa_{-n})\Bigl(G_{S,n}+\frac{1}{G_{S,n}}\Bigr)
    -2\sqrt{\kappa_{+n}\kappa_{-n}}\Bigl(G_{S,n}-\frac{1}{G_{S,n}}\Bigr)
      \cos(\theta_{+n}+\theta_{-n})\\
    &\qquad\qquad+\,4-2(\kappa_{+n}+\kappa_{-n})
  \Bigg]
\Bigg\},
\end{aligned}
\label{eq:PSD_EPR_pair}
\end{equation}
and the total photocurrent PSD is obtained by summing over all pairs:
\begin{equation}
    \bar{S}_{II}^{\mathrm{ent}}[\Omega]
    = \bar{S}_{II,0}[\Omega]
    + \sum_{n=1}^{N} \bar{S}_{II,n}^{\mathrm{ent}}[\Omega].
    \label{eq:PSD_EPR_total}
\end{equation}

\subsection{Comparison of quantum advantage}
\label{sec:comparison_quantum_advantage}

To quantify the resilience of intra-comb-line squeezing and cross-comb-line entanglement
against realistic optical losses, we evaluate the SNR advantage of
each under a localized absorber model.
We consider a flat-top comb with $M = 2N+1$ lines,
$\alpha_{S,n} \equiv \alpha_S$ and $\alpha_{L,n} \equiv \alpha_L$,
and model the sample as a single lossy line:
\begin{equation}
    \kappa_{+m} = \kappa, \qquad
    \kappa_n = 1 \quad \text{for all } n \neq +m,
\end{equation}
so that the symmetric partner at $-m$ remains perfectly transparent
($\kappa_{-m} = 1$).
As established in Eq.~\eqref{eq:SNR_kappa}, the SNR advantage
factor relative to the unsqueezed baseline reduces to the ratio of
photocurrent PSDs:
\begin{equation}
    \mathcal{A}_{\mathrm{SNR}}(G)
    \equiv
    \frac{\bar{S}_{II}[\Omega]\big|_{G=1}}{\bar{S}_{II}[\Omega]}.
    \label{eq:ASNR_def}
\end{equation}

For intra-comb-line squeezing with $\theta_n = 0$ and
uniform gains $G_{S,n} = G_S$, $G_{L,n} = G_L$,
Eq.~\eqref{eq:PSD_cross} yields the closed-form noise floor:
\begin{equation}
    \bar{S}_{II}^{\mathrm{sq}}[\Omega]
    = q_e^2
      \left[
        (2N+\kappa)\!\left(\frac{\alpha_S^2}{G_L}
                         + \frac{\alpha_L^2}{G_S}\right)
        + (1-\kappa)\,\alpha_L^2
      \right],
    \label{eq:PSD_CR_single_loss}
\end{equation}
with unsqueezed baseline ($G_S = G_L = 1$)
$q_e^2[(2N+\kappa)\alpha_S^2 + (2N+1)\alpha_L^2]$,
giving the advantage factor:
\begin{equation}
    \mathcal{A}_{\mathrm{SNR}}^{\mathrm{sq}}
    = \frac{(2N+\kappa)\alpha_S^2 + (2N+1)\alpha_L^2}
           {(2N+\kappa)\!\left(\dfrac{\alpha_S^2}{G_L}
                              + \dfrac{\alpha_L^2}{G_S}\right)
            + (1-\kappa)\,\alpha_L^2}.
    \label{eq:ASNR_CR}
\end{equation}

For EPR entanglement with uniform gains $G_{S,n} = G_S$,
$G_{L,n} = G_L$ and ideal phase alignment ($\theta_n = 0$),
the total noise is:
\begin{equation}
    \bar{S}_{II}^{\mathrm{ent}}[\Omega]
    = q_e^2
      \left[
        \frac{4N-2+(1+\sqrt{\kappa})^2}{2}
        \!\left(\frac{\alpha_S^2}{G_L}+\frac{\alpha_L^2}{G_S}\right)
        + \frac{(1-\sqrt{\kappa})^2}{2}
          \!\left(\alpha_S^2 G_L + \alpha_L^2 G_S\right)
        + (1-\kappa)\,\alpha_L^2
      \right],
    \label{eq:PSD_EPR_single_loss}
\end{equation}
with the same unsqueezed baseline
$(2N+\kappa)\alpha_S^2 + (2N+1)\alpha_L^2$ as the
intra-comb-line squeezing case, giving:
\begin{equation}
    \mathcal{A}_{\mathrm{SNR}}^{\mathrm{ent}}
    = \frac{(2N+\kappa)\alpha_S^2 + (2N+1)\alpha_L^2}
           {\dfrac{4N-2+(1+\sqrt{\kappa})^2}{2}
            \!\left(\dfrac{\alpha_S^2}{G_L}+\dfrac{\alpha_L^2}{G_S}\right)
            + \dfrac{(1-\sqrt{\kappa})^2}{2}
              \!\left(\alpha_S^2 G_L + \alpha_L^2 G_S\right)
            + (1-\kappa)\,\alpha_L^2}.
    \label{eq:ASNR_EPR}
\end{equation}

Equations~\eqref{eq:ASNR_CR} and~\eqref{eq:ASNR_EPR} together define the design space discussed in the main text: intra-comb-line squeezing is robust against localised absorption (the $\kappa$-dependent penalty enters only through the $1/G_{S/L}$ terms), whereas EPR entanglement is fragile to asymmetric loss (the anti-squeezing terms $\alpha_S^2 G_L + \alpha_L^2 G_S$ grow with the squeezing gains and with the degree of absorption asymmetry $(1-\sqrt\kappa)^2$).

\subsection{Numerical simulation of quantum-enhanced DCS}

The SNR advantage curves in Fig.~\ref{fig:DCS}(c) are computed from the closed-form noise expressions derived in Sec.~\ref{sec:dualcomb} for intra-comb-line squeezing and cross-comb-line entanglement.
A 15 dB squeezing level ($G \approx 31.6$) is applied uniformly across all comb lines, and the localized absorber model assigns a single-line absorption depth $-10\log_{10}\kappa$ to the $n = +1$ signal comb line only, with all other lines transparent ($\kappa_n = 1$ for $n \neq 1$).
Results are shown for ratios of intact-to-absorbed comb lines of $10$ and $100$ in the strong-LO approximation ($|\alpha_{L,n}| \gg |\alpha_{S,n}|$).

\end{document}